\documentclass[longauth]{aa}

\usepackage{graphicx}
\usepackage{natbib}
\usepackage{scalerel}
\usepackage{comment}
\usepackage{adjustbox}
\usepackage{longtable}
\usepackage{pdflscape}
\usepackage{fancyhdr}
\usepackage{stfloats}
\usepackage{textcomp}
\usepackage{float}
\fancypagestyle{nofooter}{
  \fancyhf{} 
}
\usepackage[table]{xcolor}
\usepackage{placeins}
\usepackage{afterpage} 

\usepackage{txfonts}
\usepackage[pdfencoding=auto,psdextra]{hyperref}
\hypersetup{
    colorlinks=true,
    linkcolor=blue,
    filecolor=magenta,      
    urlcolor=blue,
    citecolor=blue
}
\makeatletter
\renewcommand*\aa@pageof{, page \thepage{} of 23}
\makeatother

\usepackage{siunitx}

\usepackage[utf8]{inputenc}

\usepackage[switch, modulo]{lineno}

\usepackage{euclid}
\usepackage{cleveref}

\makeatletter
\let\linenumbers\relax
\makeatother

\begin{document}
%
%
   \title{Euclid Quick Data Release (Q1)}
\subtitle{The Strong Lensing Discovery Engine F -- Bright and low-redshift strong lenses}

\newcommand{\orcid}[1]{} 
\author{Euclid Collaboration: L.~R.~Ecker\orcid{0009-0005-3508-2469}\thanks{\email{ecker@usm.lmu.de}}\inst{\ref{aff1},\ref{aff2}}
\and M.~Fabricius\orcid{0000-0002-7025-6058}\inst{\ref{aff2},\ref{aff1}}
\and S.~Seitz\thanks{Deceased}\inst{\ref{aff1},\ref{aff2}}
\and R.~Saglia\orcid{0000-0003-0378-7032}\inst{\ref{aff1},\ref{aff2}}
\and N.~E.~P.~Lines\orcid{0009-0004-7751-1914}\inst{\ref{aff3}}
\and P.~Holloway\orcid{0009-0002-8896-6100}\inst{\ref{aff3}}
\and T.~Li\orcid{0009-0005-5008-0381}\inst{\ref{aff3}}
\and A.~Verma\orcid{0000-0002-0730-0781}\inst{\ref{aff4}}
\and F.~Balzer\orcid{0009-0005-6733-5432}\inst{\ref{aff2}}
\and Q.~Jin\inst{\ref{aff1}}
\and A.~Manj\'on-Garc\'ia\orcid{0000-0002-7413-8825}\inst{\ref{aff5}}
\and S.~H.~Vincken\orcid{0009-0005-7305-2359}\inst{\ref{aff6}}
\and J.~Wilde\orcid{0000-0002-4460-7379}\inst{\ref{aff7}}
\and J.~A.~Acevedo~Barroso\orcid{0000-0002-9654-1711}\inst{\ref{aff8},\ref{aff9}}
\and J.~W.~Nightingale\orcid{0000-0002-8987-7401}\inst{\ref{aff10}}
\and K.~Rojas\orcid{0000-0003-1391-6854}\inst{\ref{aff11}}
\and S.~Schuldt\orcid{0000-0003-2497-6334}\inst{\ref{aff12},\ref{aff13}}
\and M.~Walmsley\orcid{0000-0002-6408-4181}\inst{\ref{aff14},\ref{aff15}}
\and T.~E.~Collett\orcid{0000-0001-5564-3140}\inst{\ref{aff3}}
\and G.~Despali\orcid{0000-0001-6150-4112}\inst{\ref{aff16},\ref{aff17},\ref{aff18}}
\and A.~Sonnenfeld\orcid{0000-0002-6061-5977}\inst{\ref{aff19}}
\and C.~Tortora\orcid{0000-0001-7958-6531}\inst{\ref{aff20}}
\and R.~B.~Metcalf\orcid{0000-0003-3167-2574}\inst{\ref{aff16},\ref{aff17}}
\and R.~Bender\orcid{0000-0001-7179-0626}\inst{\ref{aff2},\ref{aff1}}
\and C.~Saulder\orcid{0000-0002-0408-5633}\inst{\ref{aff2},\ref{aff1}}
\and E.~Baeten\inst{\ref{aff21}}
\and C.~Cornen\orcid{0000-0002-7786-2798}\inst{\ref{aff21}}
\and D.~Delley\orcid{0000-0002-4958-7469}\inst{\ref{aff2}}
\and K.~Finner\orcid{0000-0002-4462-0709}\inst{\ref{aff22}}
\and A.~Galan\orcid{0000-0003-2547-9815}\inst{\ref{aff23},\ref{aff24}}
\and R.~Gavazzi\orcid{0000-0002-5540-6935}\inst{\ref{aff25},\ref{aff26}}
\and L.~C.~Johnson\orcid{0000-0001-6421-0953}\inst{\ref{aff27}}
\and L.~Leuzzi\orcid{0009-0006-4479-7017}\inst{\ref{aff17}}
\and C.~Macmillan\inst{\ref{aff21}}
\and P.~J.~Marshall\orcid{0000-0002-0113-5770}\inst{\ref{aff28},\ref{aff29}}
\and M.~Millon\orcid{0000-0001-7051-497X}\inst{\ref{aff30}}
\and A.~More\orcid{0000-0001-7714-7076}\inst{\ref{aff31},\ref{aff32}}
\and L.~A.~Moustakas\orcid{0000-0003-3030-2360}\inst{\ref{aff9}}
\and J.~Pearson\orcid{0000-0001-8555-8561}\inst{\ref{aff33}}
\and J.-N.~Pippert\orcid{0009-0006-9461-002X}\inst{\ref{aff2}}
\and C.~Scarlata\orcid{0000-0002-9136-8876}\inst{\ref{aff34}}
\and D.~Sluse\orcid{0000-0001-6116-2095}\inst{\ref{aff35}}
\and C.~Spiniello\orcid{0000-0002-3909-6359}\inst{\ref{aff36},\ref{aff4}}
\and T.~T.~Thai\orcid{0000-0002-8408-4816}\inst{\ref{aff37}}
\and L.~Ulivi\orcid{0009-0001-3291-5382}\inst{\ref{aff38},\ref{aff39}}
\and Han.~Wang\orcid{0000-0002-1293-5503}\inst{\ref{aff23},\ref{aff24}}
\and X.~Xu\orcid{0009-0002-0985-2309}\inst{\ref{aff40},\ref{aff41}}
\and F.~Courbin\orcid{0000-0003-0758-6510}\inst{\ref{aff7},\ref{aff42},\ref{aff43}}
\and M.~Meneghetti\orcid{0000-0003-1225-7084}\inst{\ref{aff17},\ref{aff18}}
\and N.~Aghanim\orcid{0000-0002-6688-8992}\inst{\ref{aff44}}
\and B.~Altieri\orcid{0000-0003-3936-0284}\inst{\ref{aff45}}
\and S.~Andreon\orcid{0000-0002-2041-8784}\inst{\ref{aff46}}
\and N.~Auricchio\orcid{0000-0003-4444-8651}\inst{\ref{aff17}}
\and C.~Baccigalupi\orcid{0000-0002-8211-1630}\inst{\ref{aff47},\ref{aff48},\ref{aff49},\ref{aff50}}
\and M.~Baldi\orcid{0000-0003-4145-1943}\inst{\ref{aff51},\ref{aff17},\ref{aff18}}
\and A.~Balestra\orcid{0000-0002-6967-261X}\inst{\ref{aff52}}
\and S.~Bardelli\orcid{0000-0002-8900-0298}\inst{\ref{aff17}}
\and P.~Battaglia\orcid{0000-0002-7337-5909}\inst{\ref{aff17}}
\and A.~Biviano\orcid{0000-0002-0857-0732}\inst{\ref{aff48},\ref{aff47}}
\and E.~Branchini\orcid{0000-0002-0808-6908}\inst{\ref{aff53},\ref{aff54},\ref{aff46}}
\and M.~Brescia\orcid{0000-0001-9506-5680}\inst{\ref{aff55},\ref{aff20}}
\and S.~Camera\orcid{0000-0003-3399-3574}\inst{\ref{aff56},\ref{aff57},\ref{aff58}}
\and G.~Ca\~nas-Herrera\orcid{0000-0003-2796-2149}\inst{\ref{aff59},\ref{aff60}}
\and V.~Capobianco\orcid{0000-0002-3309-7692}\inst{\ref{aff58}}
\and C.~Carbone\orcid{0000-0003-0125-3563}\inst{\ref{aff13}}
\and J.~Carretero\orcid{0000-0002-3130-0204}\inst{\ref{aff61},\ref{aff62}}
\and S.~Casas\orcid{0000-0002-4751-5138}\inst{\ref{aff63},\ref{aff64}}
\and M.~Castellano\orcid{0000-0001-9875-8263}\inst{\ref{aff65}}
\and G.~Castignani\orcid{0000-0001-6831-0687}\inst{\ref{aff17}}
\and S.~Cavuoti\orcid{0000-0002-3787-4196}\inst{\ref{aff20},\ref{aff66}}
\and K.~C.~Chambers\orcid{0000-0001-6965-7789}\inst{\ref{aff67}}
\and A.~Cimatti\inst{\ref{aff68}}
\and C.~Colodro-Conde\inst{\ref{aff69}}
\and G.~Congedo\orcid{0000-0003-2508-0046}\inst{\ref{aff59}}
\and C.~J.~Conselice\orcid{0000-0003-1949-7638}\inst{\ref{aff15}}
\and L.~Conversi\orcid{0000-0002-6710-8476}\inst{\ref{aff70},\ref{aff45}}
\and Y.~Copin\orcid{0000-0002-5317-7518}\inst{\ref{aff71}}
\and A.~Costille\inst{\ref{aff25}}
\and H.~M.~Courtois\orcid{0000-0003-0509-1776}\inst{\ref{aff72}}
\and M.~Cropper\orcid{0000-0003-4571-9468}\inst{\ref{aff73}}
\and A.~Da~Silva\orcid{0000-0002-6385-1609}\inst{\ref{aff74},\ref{aff75}}
\and H.~Degaudenzi\orcid{0000-0002-5887-6799}\inst{\ref{aff76}}
\and G.~De~Lucia\orcid{0000-0002-6220-9104}\inst{\ref{aff48}}
\and C.~Dolding\orcid{0009-0003-7199-6108}\inst{\ref{aff73}}
\and H.~Dole\orcid{0000-0002-9767-3839}\inst{\ref{aff44}}
\and F.~Dubath\orcid{0000-0002-6533-2810}\inst{\ref{aff76}}
\and X.~Dupac\inst{\ref{aff45}}
\and S.~Dusini\orcid{0000-0002-1128-0664}\inst{\ref{aff77}}
\and A.~Ealet\orcid{0000-0003-3070-014X}\inst{\ref{aff71}}
\and S.~Escoffier\orcid{0000-0002-2847-7498}\inst{\ref{aff78}}
\and M.~Farina\orcid{0000-0002-3089-7846}\inst{\ref{aff79}}
\and R.~Farinelli\inst{\ref{aff17}}
\and F.~Faustini\orcid{0000-0001-6274-5145}\inst{\ref{aff65},\ref{aff80}}
\and S.~Ferriol\inst{\ref{aff71}}
\and F.~Finelli\orcid{0000-0002-6694-3269}\inst{\ref{aff17},\ref{aff81}}
\and P.~Fosalba\orcid{0000-0002-1510-5214}\inst{\ref{aff82},\ref{aff83}}
\and S.~Fotopoulou\orcid{0000-0002-9686-254X}\inst{\ref{aff84}}
\and M.~Frailis\orcid{0000-0002-7400-2135}\inst{\ref{aff48}}
\and E.~Franceschi\orcid{0000-0002-0585-6591}\inst{\ref{aff17}}
\and M.~Fumana\orcid{0000-0001-6787-5950}\inst{\ref{aff13}}
\and S.~Galeotta\orcid{0000-0002-3748-5115}\inst{\ref{aff48}}
\and K.~George\orcid{0000-0002-1734-8455}\inst{\ref{aff85}}
\and W.~Gillard\orcid{0000-0003-4744-9748}\inst{\ref{aff78}}
\and B.~Gillis\orcid{0000-0002-4478-1270}\inst{\ref{aff59}}
\and C.~Giocoli\orcid{0000-0002-9590-7961}\inst{\ref{aff17},\ref{aff18}}
\and P.~G\'omez-Alvarez\orcid{0000-0002-8594-5358}\inst{\ref{aff86},\ref{aff45}}
\and J.~Gracia-Carpio\inst{\ref{aff2}}
\and A.~Grazian\orcid{0000-0002-5688-0663}\inst{\ref{aff52}}
\and F.~Grupp\inst{\ref{aff2},\ref{aff1}}
\and L.~Guzzo\orcid{0000-0001-8264-5192}\inst{\ref{aff12},\ref{aff46},\ref{aff87}}
\and S.~V.~H.~Haugan\orcid{0000-0001-9648-7260}\inst{\ref{aff88}}
\and H.~Hoekstra\orcid{0000-0002-0641-3231}\inst{\ref{aff60}}
\and W.~Holmes\inst{\ref{aff9}}
\and F.~Hormuth\inst{\ref{aff89}}
\and A.~Hornstrup\orcid{0000-0002-3363-0936}\inst{\ref{aff90},\ref{aff91}}
\and K.~Jahnke\orcid{0000-0003-3804-2137}\inst{\ref{aff92}}
\and M.~Jhabvala\inst{\ref{aff93}}
\and B.~Joachimi\orcid{0000-0001-7494-1303}\inst{\ref{aff94}}
\and E.~Keih\"anen\orcid{0000-0003-1804-7715}\inst{\ref{aff95}}
\and S.~Kermiche\orcid{0000-0002-0302-5735}\inst{\ref{aff78}}
\and A.~Kiessling\orcid{0000-0002-2590-1273}\inst{\ref{aff9}}
\and B.~Kubik\orcid{0009-0006-5823-4880}\inst{\ref{aff71}}
\and M.~K\"ummel\orcid{0000-0003-2791-2117}\inst{\ref{aff1}}
\and M.~Kunz\orcid{0000-0002-3052-7394}\inst{\ref{aff30}}
\and H.~Kurki-Suonio\orcid{0000-0002-4618-3063}\inst{\ref{aff96},\ref{aff97}}
\and A.~M.~C.~Le~Brun\orcid{0000-0002-0936-4594}\inst{\ref{aff98}}
\and D.~Le~Mignant\orcid{0000-0002-5339-5515}\inst{\ref{aff25}}
\and S.~Ligori\orcid{0000-0003-4172-4606}\inst{\ref{aff58}}
\and P.~B.~Lilje\orcid{0000-0003-4324-7794}\inst{\ref{aff88}}
\and V.~Lindholm\orcid{0000-0003-2317-5471}\inst{\ref{aff96},\ref{aff97}}
\and I.~Lloro\orcid{0000-0001-5966-1434}\inst{\ref{aff99}}
\and G.~Mainetti\orcid{0000-0003-2384-2377}\inst{\ref{aff100}}
\and D.~Maino\inst{\ref{aff12},\ref{aff13},\ref{aff87}}
\and E.~Maiorano\orcid{0000-0003-2593-4355}\inst{\ref{aff17}}
\and O.~Mansutti\orcid{0000-0001-5758-4658}\inst{\ref{aff48}}
\and S.~Marcin\inst{\ref{aff11}}
\and O.~Marggraf\orcid{0000-0001-7242-3852}\inst{\ref{aff101}}
\and M.~Martinelli\orcid{0000-0002-6943-7732}\inst{\ref{aff65},\ref{aff102}}
\and N.~Martinet\orcid{0000-0003-2786-7790}\inst{\ref{aff25}}
\and F.~Marulli\orcid{0000-0002-8850-0303}\inst{\ref{aff16},\ref{aff17},\ref{aff18}}
\and R.~J.~Massey\orcid{0000-0002-6085-3780}\inst{\ref{aff103}}
\and E.~Medinaceli\orcid{0000-0002-4040-7783}\inst{\ref{aff17}}
\and S.~Mei\orcid{0000-0002-2849-559X}\inst{\ref{aff104},\ref{aff105}}
\and Y.~Mellier$^{\star\star}$\inst{\ref{aff106},\ref{aff26}}
\and E.~Merlin\orcid{0000-0001-6870-8900}\inst{\ref{aff65}}
\and G.~Meylan\inst{\ref{aff8}}
\and A.~Mora\orcid{0000-0002-1922-8529}\inst{\ref{aff107}}
\and M.~Moresco\orcid{0000-0002-7616-7136}\inst{\ref{aff16},\ref{aff17}}
\and L.~Moscardini\orcid{0000-0002-3473-6716}\inst{\ref{aff16},\ref{aff17},\ref{aff18}}
\and R.~Nakajima\orcid{0009-0009-1213-7040}\inst{\ref{aff101}}
\and C.~Neissner\orcid{0000-0001-8524-4968}\inst{\ref{aff108},\ref{aff62}}
\and R.~C.~Nichol\orcid{0000-0003-0939-6518}\inst{\ref{aff109}}
\and S.-M.~Niemi\orcid{0009-0005-0247-0086}\inst{\ref{aff110}}
\and C.~Padilla\orcid{0000-0001-7951-0166}\inst{\ref{aff108}}
\and S.~Paltani\orcid{0000-0002-8108-9179}\inst{\ref{aff76}}
\and F.~Pasian\orcid{0000-0002-4869-3227}\inst{\ref{aff48}}
\and K.~Pedersen\inst{\ref{aff111}}
\and W.~J.~Percival\orcid{0000-0002-0644-5727}\inst{\ref{aff112},\ref{aff113},\ref{aff114}}
\and V.~Pettorino\orcid{0000-0002-4203-9320}\inst{\ref{aff110}}
\and S.~Pires\orcid{0000-0002-0249-2104}\inst{\ref{aff115}}
\and G.~Polenta\orcid{0000-0003-4067-9196}\inst{\ref{aff80}}
\and M.~Poncet\inst{\ref{aff116}}
\and L.~Pozzetti\orcid{0000-0001-7085-0412}\inst{\ref{aff17}}
\and F.~Raison\orcid{0000-0002-7819-6918}\inst{\ref{aff2}}
\and A.~Renzi\orcid{0000-0001-9856-1970}\inst{\ref{aff117},\ref{aff77}}
\and J.~Rhodes\orcid{0000-0002-4485-8549}\inst{\ref{aff9}}
\and G.~Riccio\inst{\ref{aff20}}
\and H.-W.~Rix\orcid{0000-0003-4996-9069}\inst{\ref{aff92}}
\and E.~Romelli\orcid{0000-0003-3069-9222}\inst{\ref{aff48}}
\and M.~Roncarelli\orcid{0000-0001-9587-7822}\inst{\ref{aff17}}
\and E.~Rossetti\orcid{0000-0003-0238-4047}\inst{\ref{aff51}}
\and Z.~Sakr\orcid{0000-0002-4823-3757}\inst{\ref{aff118},\ref{aff119},\ref{aff120}}
\and A.~G.~S\'anchez\orcid{0000-0003-1198-831X}\inst{\ref{aff2}}
\and D.~Sapone\orcid{0000-0001-7089-4503}\inst{\ref{aff121}}
\and B.~Sartoris\orcid{0000-0003-1337-5269}\inst{\ref{aff1},\ref{aff48}}
\and P.~Schneider\orcid{0000-0001-8561-2679}\inst{\ref{aff101}}
\and T.~Schrabback\orcid{0000-0002-6987-7834}\inst{\ref{aff122}}
\and A.~Secroun\orcid{0000-0003-0505-3710}\inst{\ref{aff78}}
\and G.~Seidel\orcid{0000-0003-2907-353X}\inst{\ref{aff92}}
\and S.~Serrano\orcid{0000-0002-0211-2861}\inst{\ref{aff82},\ref{aff123},\ref{aff83}}
\and P.~Simon\inst{\ref{aff101}}
\and C.~Sirignano\orcid{0000-0002-0995-7146}\inst{\ref{aff117},\ref{aff77}}
\and G.~Sirri\orcid{0000-0003-2626-2853}\inst{\ref{aff18}}
\and L.~Stanco\orcid{0000-0002-9706-5104}\inst{\ref{aff77}}
\and J.~Steinwagner\orcid{0000-0001-7443-1047}\inst{\ref{aff2}}
\and P.~Tallada-Cresp\'{i}\orcid{0000-0002-1336-8328}\inst{\ref{aff61},\ref{aff62}}
\and A.~N.~Taylor\inst{\ref{aff59}}
\and H.~I.~Teplitz\orcid{0000-0002-7064-5424}\inst{\ref{aff124}}
\and I.~Tereno\orcid{0000-0002-4537-6218}\inst{\ref{aff74},\ref{aff125}}
\and N.~Tessore\orcid{0000-0002-9696-7931}\inst{\ref{aff73}}
\and S.~Toft\orcid{0000-0003-3631-7176}\inst{\ref{aff126},\ref{aff127}}
\and R.~Toledo-Moreo\orcid{0000-0002-2997-4859}\inst{\ref{aff128}}
\and F.~Torradeflot\orcid{0000-0003-1160-1517}\inst{\ref{aff62},\ref{aff61}}
\and I.~Tutusaus\orcid{0000-0002-3199-0399}\inst{\ref{aff83},\ref{aff82},\ref{aff119}}
\and L.~Valenziano\orcid{0000-0002-1170-0104}\inst{\ref{aff17},\ref{aff81}}
\and J.~Valiviita\orcid{0000-0001-6225-3693}\inst{\ref{aff96},\ref{aff97}}
\and T.~Vassallo\orcid{0000-0001-6512-6358}\inst{\ref{aff48}}
\and Y.~Wang\orcid{0000-0002-4749-2984}\inst{\ref{aff22}}
\and J.~Weller\orcid{0000-0002-8282-2010}\inst{\ref{aff1},\ref{aff2}}
\and A.~Zacchei\orcid{0000-0003-0396-1192}\inst{\ref{aff48},\ref{aff47}}
\and G.~Zamorani\orcid{0000-0002-2318-301X}\inst{\ref{aff17}}
\and F.~M.~Zerbi\inst{\ref{aff46}}
\and E.~Zucca\orcid{0000-0002-5845-8132}\inst{\ref{aff17}}
\and M.~Ballardini\orcid{0000-0003-4481-3559}\inst{\ref{aff129},\ref{aff130},\ref{aff17}}
\and M.~Bolzonella\orcid{0000-0003-3278-4607}\inst{\ref{aff17}}
\and E.~Bozzo\orcid{0000-0002-8201-1525}\inst{\ref{aff76}}
\and C.~Burigana\orcid{0000-0002-3005-5796}\inst{\ref{aff131},\ref{aff81}}
\and R.~Cabanac\orcid{0000-0001-6679-2600}\inst{\ref{aff119}}
\and A.~Cappi\inst{\ref{aff132},\ref{aff17}}
\and T.~Castro\orcid{0000-0002-6292-3228}\inst{\ref{aff48},\ref{aff49},\ref{aff47},\ref{aff133}}
\and B.~Cl\'ement\orcid{0000-0002-7966-3661}\inst{\ref{aff8},\ref{aff134}}
\and J.~A.~Escartin~Vigo\inst{\ref{aff2}}
\and L.~Gabarra\orcid{0000-0002-8486-8856}\inst{\ref{aff4}}
\and J.~Garc\'ia-Bellido\orcid{0000-0002-9370-8360}\inst{\ref{aff135}}
\and V.~Gautard\inst{\ref{aff136}}
\and S.~Hemmati\orcid{0000-0003-2226-5395}\inst{\ref{aff22}}
\and M.~Huertas-Company\orcid{0000-0002-1416-8483}\inst{\ref{aff69},\ref{aff137},\ref{aff138}}
\and J.~Macias-Perez\orcid{0000-0002-5385-2763}\inst{\ref{aff139}}
\and R.~Maoli\orcid{0000-0002-6065-3025}\inst{\ref{aff140},\ref{aff65}}
\and J.~Mart\'{i}n-Fleitas\orcid{0000-0002-8594-569X}\inst{\ref{aff141}}
\and M.~Maturi\orcid{0000-0002-3517-2422}\inst{\ref{aff118},\ref{aff142}}
\and N.~Mauri\orcid{0000-0001-8196-1548}\inst{\ref{aff68},\ref{aff18}}
\and P.~Monaco\orcid{0000-0003-2083-7564}\inst{\ref{aff143},\ref{aff48},\ref{aff49},\ref{aff47}}
\and A.~Pezzotta\orcid{0000-0003-0726-2268}\inst{\ref{aff46}}
\and M.~P\"ontinen\orcid{0000-0001-5442-2530}\inst{\ref{aff96}}
\and C.~Porciani\orcid{0000-0002-7797-2508}\inst{\ref{aff101}}
\and I.~Risso\orcid{0000-0003-2525-7761}\inst{\ref{aff46},\ref{aff54}}
\and V.~Scottez\orcid{0009-0008-3864-940X}\inst{\ref{aff106},\ref{aff144}}
\and M.~Sereno\orcid{0000-0003-0302-0325}\inst{\ref{aff17},\ref{aff18}}
\and M.~Tenti\orcid{0000-0002-4254-5901}\inst{\ref{aff18}}
\and M.~Tucci\inst{\ref{aff76}}
\and M.~Viel\orcid{0000-0002-2642-5707}\inst{\ref{aff47},\ref{aff48},\ref{aff50},\ref{aff49},\ref{aff133}}
\and M.~Wiesmann\orcid{0009-0000-8199-5860}\inst{\ref{aff88}}
\and Y.~Akrami\orcid{0000-0002-2407-7956}\inst{\ref{aff135},\ref{aff145}}
\and I.~T.~Andika\orcid{0000-0001-6102-9526}\inst{\ref{aff85},\ref{aff24}}
\and G.~Angora\orcid{0000-0002-0316-6562}\inst{\ref{aff20},\ref{aff129}}
\and S.~Anselmi\orcid{0000-0002-3579-9583}\inst{\ref{aff77},\ref{aff117},\ref{aff146}}
\and M.~Archidiacono\orcid{0000-0003-4952-9012}\inst{\ref{aff12},\ref{aff87}}
\and F.~Atrio-Barandela\orcid{0000-0002-2130-2513}\inst{\ref{aff147}}
\and L.~Bazzanini\orcid{0000-0003-0727-0137}\inst{\ref{aff129},\ref{aff17}}
\and P.~Bergamini\orcid{0000-0003-1383-9414}\inst{\ref{aff17}}
\and D.~Bertacca\orcid{0000-0002-2490-7139}\inst{\ref{aff117},\ref{aff52},\ref{aff77}}
\and M.~Bethermin\orcid{0000-0002-3915-2015}\inst{\ref{aff148}}
\and F.~Beutler\orcid{0000-0003-0467-5438}\inst{\ref{aff59}}
\and A.~Blanchard\orcid{0000-0001-8555-9003}\inst{\ref{aff119}}
\and L.~Blot\orcid{0000-0002-9622-7167}\inst{\ref{aff149},\ref{aff98}}
\and M.~Bonici\orcid{0000-0002-8430-126X}\inst{\ref{aff112},\ref{aff13}}
\and S.~Borgani\orcid{0000-0001-6151-6439}\inst{\ref{aff143},\ref{aff47},\ref{aff48},\ref{aff49},\ref{aff133}}
\and M.~L.~Brown\orcid{0000-0002-0370-8077}\inst{\ref{aff15}}
\and S.~Bruton\orcid{0000-0002-6503-5218}\inst{\ref{aff150}}
\and A.~Calabro\orcid{0000-0003-2536-1614}\inst{\ref{aff65}}
\and B.~Camacho~Quevedo\orcid{0000-0002-8789-4232}\inst{\ref{aff47},\ref{aff50},\ref{aff48}}
\and F.~Caro\inst{\ref{aff65}}
\and C.~S.~Carvalho\inst{\ref{aff125}}
\and Y.~Charles\inst{\ref{aff25}}
\and F.~Cogato\orcid{0000-0003-4632-6113}\inst{\ref{aff16},\ref{aff17}}
\and S.~Conseil\orcid{0000-0002-3657-4191}\inst{\ref{aff71}}
\and A.~R.~Cooray\orcid{0000-0002-3892-0190}\inst{\ref{aff40}}
\and O.~Cucciati\orcid{0000-0002-9336-7551}\inst{\ref{aff17}}
\and S.~Davini\orcid{0000-0003-3269-1718}\inst{\ref{aff54}}
\and F.~De~Paolis\orcid{0000-0001-6460-7563}\inst{\ref{aff151},\ref{aff152},\ref{aff153}}
\and G.~Desprez\orcid{0000-0001-8325-1742}\inst{\ref{aff154}}
\and A.~D\'iaz-S\'anchez\orcid{0000-0003-0748-4768}\inst{\ref{aff5}}
\and S.~Di~Domizio\orcid{0000-0003-2863-5895}\inst{\ref{aff53},\ref{aff54}}
\and J.~M.~Diego\orcid{0000-0001-9065-3926}\inst{\ref{aff155}}
\and P.-A.~Duc\orcid{0000-0003-3343-6284}\inst{\ref{aff148}}
\and V.~Duret\orcid{0009-0009-0383-4960}\inst{\ref{aff78}}
\and M.~Y.~Elkhashab\orcid{0000-0001-9306-2603}\inst{\ref{aff48},\ref{aff49},\ref{aff143},\ref{aff47}}
\and A.~Enia\orcid{0000-0002-0200-2857}\inst{\ref{aff17}}
\and Y.~Fang\orcid{0000-0002-0334-6950}\inst{\ref{aff1}}
\and A.~Finoguenov\orcid{0000-0002-4606-5403}\inst{\ref{aff96}}
\and A.~Fontana\orcid{0000-0003-3820-2823}\inst{\ref{aff65}}
\and A.~Franco\orcid{0000-0002-4761-366X}\inst{\ref{aff152},\ref{aff151},\ref{aff153}}
\and K.~Ganga\orcid{0000-0001-8159-8208}\inst{\ref{aff104}}
\and T.~Gasparetto\orcid{0000-0002-7913-4866}\inst{\ref{aff65}}
\and E.~Gaztanaga\orcid{0000-0001-9632-0815}\inst{\ref{aff83},\ref{aff82},\ref{aff3}}
\and F.~Giacomini\orcid{0000-0002-3129-2814}\inst{\ref{aff18}}
\and F.~Gianotti\orcid{0000-0003-4666-119X}\inst{\ref{aff17}}
\and G.~Gozaliasl\orcid{0000-0002-0236-919X}\inst{\ref{aff156},\ref{aff96}}
\and A.~Gruppuso\orcid{0000-0001-9272-5292}\inst{\ref{aff17},\ref{aff18}}
\and M.~Guidi\orcid{0000-0001-9408-1101}\inst{\ref{aff51},\ref{aff17}}
\and C.~M.~Gutierrez\orcid{0000-0001-7854-783X}\inst{\ref{aff157}}
\and A.~Hall\orcid{0000-0002-3139-8651}\inst{\ref{aff59}}
\and H.~Hildebrandt\orcid{0000-0002-9814-3338}\inst{\ref{aff158}}
\and J.~Hjorth\orcid{0000-0002-4571-2306}\inst{\ref{aff111}}
\and L.~K.~Hunt\orcid{0000-0001-9162-2371}\inst{\ref{aff39}}
\and J.~J.~E.~Kajava\orcid{0000-0002-3010-8333}\inst{\ref{aff159},\ref{aff160},\ref{aff161}}
\and Y.~Kang\orcid{0009-0000-8588-7250}\inst{\ref{aff76}}
\and V.~Kansal\orcid{0000-0002-4008-6078}\inst{\ref{aff162},\ref{aff163}}
\and D.~Karagiannis\orcid{0000-0002-4927-0816}\inst{\ref{aff129},\ref{aff164}}
\and K.~Kiiveri\inst{\ref{aff95}}
\and J.~Kim\orcid{0000-0003-2776-2761}\inst{\ref{aff4}}
\and C.~C.~Kirkpatrick\inst{\ref{aff95}}
\and S.~Kruk\orcid{0000-0001-8010-8879}\inst{\ref{aff45}}
\and M.~Lattanzi\orcid{0000-0003-1059-2532}\inst{\ref{aff130}}
\and L.~Legrand\orcid{0000-0003-0610-5252}\inst{\ref{aff165},\ref{aff166}}
\and F.~Lepori\orcid{0009-0000-5061-7138}\inst{\ref{aff167}}
\and G.~Leroy\orcid{0009-0004-2523-4425}\inst{\ref{aff168},\ref{aff103}}
\and G.~F.~Lesci\orcid{0000-0002-4607-2830}\inst{\ref{aff16},\ref{aff17}}
\and J.~Lesgourgues\orcid{0000-0001-7627-353X}\inst{\ref{aff63}}
\and T.~I.~Liaudat\orcid{0000-0002-9104-314X}\inst{\ref{aff169}}
\and A.~Loureiro\orcid{0000-0002-4371-0876}\inst{\ref{aff170},\ref{aff171}}
\and M.~Magliocchetti\orcid{0000-0001-9158-4838}\inst{\ref{aff79}}
\and F.~Mannucci\orcid{0000-0002-4803-2381}\inst{\ref{aff39}}
\and C.~J.~A.~P.~Martins\orcid{0000-0002-4886-9261}\inst{\ref{aff172},\ref{aff173}}
\and L.~Maurin\orcid{0000-0002-8406-0857}\inst{\ref{aff44}}
\and M.~Miluzio\inst{\ref{aff45},\ref{aff174}}
\and C.~Moretti\orcid{0000-0003-3314-8936}\inst{\ref{aff48},\ref{aff47},\ref{aff49}}
\and G.~Morgante\inst{\ref{aff17}}
\and K.~Naidoo\orcid{0000-0002-9182-1802}\inst{\ref{aff3},\ref{aff92}}
\and P.~Natoli\orcid{0000-0003-0126-9100}\inst{\ref{aff129},\ref{aff130}}
\and A.~Navarro-Alsina\orcid{0000-0002-3173-2592}\inst{\ref{aff101}}
\and S.~Nesseris\orcid{0000-0002-0567-0324}\inst{\ref{aff135}}
\and D.~Paoletti\orcid{0000-0003-4761-6147}\inst{\ref{aff17},\ref{aff81}}
\and F.~Passalacqua\orcid{0000-0002-8606-4093}\inst{\ref{aff117},\ref{aff77}}
\and K.~Paterson\orcid{0000-0001-8340-3486}\inst{\ref{aff92}}
\and L.~Patrizii\inst{\ref{aff18}}
\and A.~Pisani\orcid{0000-0002-6146-4437}\inst{\ref{aff78}}
\and D.~Potter\orcid{0000-0002-0757-5195}\inst{\ref{aff167}}
\and G.~W.~Pratt\inst{\ref{aff115}}
\and S.~Quai\orcid{0000-0002-0449-8163}\inst{\ref{aff16},\ref{aff17}}
\and M.~Radovich\orcid{0000-0002-3585-866X}\inst{\ref{aff52}}
\and G.~Rodighiero\orcid{0000-0002-9415-2296}\inst{\ref{aff117},\ref{aff52}}
\and W.~Roster\orcid{0000-0002-9149-6528}\inst{\ref{aff2}}
\and S.~Sacquegna\orcid{0000-0002-8433-6630}\inst{\ref{aff175}}
\and M.~Sahl\'en\orcid{0000-0003-0973-4804}\inst{\ref{aff176}}
\and D.~B.~Sanders\orcid{0000-0002-1233-9998}\inst{\ref{aff67}}
\and E.~Sarpa\orcid{0000-0002-1256-655X}\inst{\ref{aff50},\ref{aff133},\ref{aff49}}
\and A.~Schneider\orcid{0000-0001-7055-8104}\inst{\ref{aff167}}
\and D.~Sciotti\orcid{0009-0008-4519-2620}\inst{\ref{aff65},\ref{aff102}}
\and E.~Sellentin\inst{\ref{aff177},\ref{aff60}}
\and L.~C.~Smith\orcid{0000-0002-3259-2771}\inst{\ref{aff178}}
\and J.~G.~Sorce\orcid{0000-0002-2307-2432}\inst{\ref{aff179},\ref{aff44}}
\and K.~Tanidis\orcid{0000-0001-9843-5130}\inst{\ref{aff4}}
\and C.~Tao\orcid{0000-0001-7961-8177}\inst{\ref{aff78}}
\and F.~Tarsitano\orcid{0000-0002-5919-0238}\inst{\ref{aff180},\ref{aff76}}
\and G.~Testera\inst{\ref{aff54}}
\and R.~Teyssier\orcid{0000-0001-7689-0933}\inst{\ref{aff181}}
\and S.~Tosi\orcid{0000-0002-7275-9193}\inst{\ref{aff53},\ref{aff54},\ref{aff46}}
\and A.~Troja\orcid{0000-0003-0239-4595}\inst{\ref{aff117},\ref{aff77}}
\and A.~Venhola\orcid{0000-0001-6071-4564}\inst{\ref{aff182}}
\and D.~Vergani\orcid{0000-0003-0898-2216}\inst{\ref{aff17}}
\and G.~Vernardos\orcid{0000-0001-8554-7248}\inst{\ref{aff183},\ref{aff184}}
\and G.~Verza\orcid{0000-0002-1886-8348}\inst{\ref{aff185},\ref{aff186}}
\and P.~Vielzeuf\orcid{0000-0003-2035-9339}\inst{\ref{aff78}}
\and S.~Vinciguerra\orcid{0009-0005-4018-3184}\inst{\ref{aff25}}
\and N.~A.~Walton\orcid{0000-0003-3983-8778}\inst{\ref{aff178}}
\and A.~H.~Wright\orcid{0000-0001-7363-7932}\inst{\ref{aff158}}}
										   
\institute{Universit\"ats-Sternwarte M\"unchen, Fakult\"at f\"ur Physik, Ludwig-Maximilians-Universit\"at M\"unchen, Scheinerstr.~1, 81679 M\"unchen, Germany\label{aff1}
\and
Max Planck Institute for Extraterrestrial Physics, Giessenbachstr. 1, 85748 Garching, Germany\label{aff2}
\and
Institute of Cosmology and Gravitation, University of Portsmouth, Portsmouth PO1 3FX, UK\label{aff3}
\and
Department of Physics, Oxford University, Keble Road, Oxford OX1 3RH, UK\label{aff4}
\and
Departamento F\'isica Aplicada, Universidad Polit\'ecnica de Cartagena, Campus Muralla del Mar, 30202 Cartagena, Murcia, Spain\label{aff5}
\and
University of Applied Sciences and Arts of Northwestern Switzerland, School of Engineering, 5210 Windisch, Switzerland\label{aff6}
\and
Institut de Ci\`{e}ncies del Cosmos (ICCUB), Universitat de Barcelona (IEEC-UB), Mart\'{i} i Franqu\`{e}s 1, 08028 Barcelona, Spain\label{aff7}
\and
Institute of Physics, Laboratory of Astrophysics, Ecole Polytechnique F\'ed\'erale de Lausanne (EPFL), Observatoire de Sauverny, 1290 Versoix, Switzerland\label{aff8}
\and
Jet Propulsion Laboratory, California Institute of Technology, 4800 Oak Grove Drive, Pasadena, CA, 91109, USA\label{aff9}
\and
School of Mathematics, Statistics and Physics, Newcastle University, Herschel Building, Newcastle-upon-Tyne, NE1 7RU, UK\label{aff10}
\and
University of Applied Sciences and Arts of Northwestern Switzerland, School of Computer Science, 5210 Windisch, Switzerland\label{aff11}
\and
Dipartimento di Fisica "Aldo Pontremoli", Universit\`a degli Studi di Milano, Via Celoria 16, 20133 Milano, Italy\label{aff12}
\and
INAF-IASF Milano, Via Alfonso Corti 12, 20133 Milano, Italy\label{aff13}
\and
David A. Dunlap Department of Astronomy \& Astrophysics, University of Toronto, 50 St George Street, Toronto, Ontario M5S 3H4, Canada\label{aff14}
\and
Jodrell Bank Centre for Astrophysics, Department of Physics and Astronomy, University of Manchester, Oxford Road, Manchester M13 9PL, UK\label{aff15}
\and
Dipartimento di Fisica e Astronomia "Augusto Righi" - Alma Mater Studiorum Universit\`a di Bologna, via Piero Gobetti 93/2, 40129 Bologna, Italy\label{aff16}
\and
INAF-Osservatorio di Astrofisica e Scienza dello Spazio di Bologna, Via Piero Gobetti 93/3, 40129 Bologna, Italy\label{aff17}
\and
INFN-Sezione di Bologna, Viale Berti Pichat 6/2, 40127 Bologna, Italy\label{aff18}
\and
Department of Astronomy, School of Physics and Astronomy, Shanghai Jiao Tong University, Shanghai 200240, China\label{aff19}
\and
INAF-Osservatorio Astronomico di Capodimonte, Via Moiariello 16, 80131 Napoli, Italy\label{aff20}
\and
Citizen Scientist, Zooniverse c/o University of Oxford,  Keble Road, Oxford OX1 3RH, UK\label{aff21}
\and
Caltech/IPAC, 1200 E. California Blvd., Pasadena, CA 91125, USA\label{aff22}
\and
Max-Planck-Institut f\"ur Astrophysik, Karl-Schwarzschild-Str.~1, 85748 Garching, Germany\label{aff23}
\and
Technical University of Munich, TUM School of Natural Sciences, Physics Department, James-Franck-Str.~1, 85748 Garching, Germany\label{aff24}
\and
Aix-Marseille Universit\'e, CNRS, CNES, LAM, Marseille, France\label{aff25}
\and
Institut d'Astrophysique de Paris, UMR 7095, CNRS, and Sorbonne Universit\'e, 98 bis boulevard Arago, 75014 Paris, France\label{aff26}
\and
Center for Interdisciplinary Exploration and Research in Astrophysics (CIERA) and Department of Physics and Astronomy, Northwestern University, 1800 Sherman Ave., Evanston, IL 60201, USA\label{aff27}
\and
Kavli Institute for Particle Astrophysics \& Cosmology (KIPAC), Stanford University, Stanford, CA 94305, USA\label{aff28}
\and
SLAC National Accelerator Laboratory, 2575 Sand Hill Road, Menlo Park, CA 94025, USA\label{aff29}
\and
Universit\'e de Gen\`eve, D\'epartement de Physique Th\'eorique and Centre for Astroparticle Physics, 24 quai Ernest-Ansermet, CH-1211 Gen\`eve 4, Switzerland\label{aff30}
\and
The Inter-University Centre for Astronomy and Astrophysics, Post Bag 4, Ganeshkhind, Pune 411007, India\label{aff31}
\and
Kavli Institute for the Physics and Mathematics of the Universe (WPI), University of Tokyo, Kashiwa, Chiba 277-8583, Japan\label{aff32}
\and
School of Physical Sciences, The Open University, Milton Keynes, MK7 6AA, UK\label{aff33}
\and
Minnesota Institute for Astrophysics, University of Minnesota, 116 Church St SE, Minneapolis, MN 55455, USA\label{aff34}
\and
STAR Institute, University of Li{\`e}ge, Quartier Agora, All\'ee du six Ao\^ut 19c, 4000 Li\`ege, Belgium\label{aff35}
\and
European Southern Observatory, Karl-Schwarzschild-Str.~2, 85748 Garching, Germany\label{aff36}
\and
National Astronomical Observatory of Japan, 2-21-1 Osawa, Mitaka, Tokyo 181-8588, Japan\label{aff37}
\and
Centro de Astrobiolog\'ia (CAB), CSIC--INTA, Cra. de Ajalvir Km.~4, 28850 -- Torrej\'on de Ardoz, Madrid, Spain\label{aff38}
\and
INAF-Osservatorio Astrofisico di Arcetri, Largo E. Fermi 5, 50125, Firenze, Italy\label{aff39}
\and
Department of Physics \& Astronomy, University of California Irvine, Irvine CA 92697, USA\label{aff40}
\and
University of Southern California, 3551 Trousdale Parkway, Los Angeles, CA 90089, USA\label{aff41}
\and
Instituci\'o Catalana de Recerca i Estudis Avan\c{c}ats (ICREA), Passeig de Llu\'{\i}s Companys 23, 08010 Barcelona, Spain\label{aff42}
\and
Institut de Ciencies de l'Espai (IEEC-CSIC), Campus UAB, Carrer de Can Magrans, s/n Cerdanyola del Vall\'es, 08193 Barcelona, Spain\label{aff43}
\and
Universit\'e Paris-Saclay, CNRS, Institut d'astrophysique spatiale, 91405, Orsay, France\label{aff44}
\and
ESAC/ESA, Camino Bajo del Castillo, s/n., Urb. Villafranca del Castillo, 28692 Villanueva de la Ca\~nada, Madrid, Spain\label{aff45}
\and
INAF-Osservatorio Astronomico di Brera, Via Brera 28, 20122 Milano, Italy\label{aff46}
\and
IFPU, Institute for Fundamental Physics of the Universe, via Beirut 2, 34151 Trieste, Italy\label{aff47}
\and
INAF-Osservatorio Astronomico di Trieste, Via G. B. Tiepolo 11, 34143 Trieste, Italy\label{aff48}
\and
INFN, Sezione di Trieste, Via Valerio 2, 34127 Trieste TS, Italy\label{aff49}
\and
SISSA, International School for Advanced Studies, Via Bonomea 265, 34136 Trieste TS, Italy\label{aff50}
\and
Dipartimento di Fisica e Astronomia, Universit\`a di Bologna, Via Gobetti 93/2, 40129 Bologna, Italy\label{aff51}
\and
INAF-Osservatorio Astronomico di Padova, Via dell'Osservatorio 5, 35122 Padova, Italy\label{aff52}
\and
Dipartimento di Fisica, Universit\`a di Genova, Via Dodecaneso 33, 16146, Genova, Italy\label{aff53}
\and
INFN-Sezione di Genova, Via Dodecaneso 33, 16146, Genova, Italy\label{aff54}
\and
Department of Physics "E. Pancini", University Federico II, Via Cinthia 6, 80126, Napoli, Italy\label{aff55}
\and
Dipartimento di Fisica, Universit\`a degli Studi di Torino, Via P. Giuria 1, 10125 Torino, Italy\label{aff56}
\and
INFN-Sezione di Torino, Via P. Giuria 1, 10125 Torino, Italy\label{aff57}
\and
INAF-Osservatorio Astrofisico di Torino, Via Osservatorio 20, 10025 Pino Torinese (TO), Italy\label{aff58}
\and
Institute for Astronomy, University of Edinburgh, Royal Observatory, Blackford Hill, Edinburgh EH9 3HJ, UK\label{aff59}
\and
Leiden Observatory, Leiden University, Einsteinweg 55, 2333 CC Leiden, The Netherlands\label{aff60}
\and
Centro de Investigaciones Energ\'eticas, Medioambientales y Tecnol\'ogicas (CIEMAT), Avenida Complutense 40, 28040 Madrid, Spain\label{aff61}
\and
Port d'Informaci\'{o} Cient\'{i}fica, Campus UAB, C. Albareda s/n, 08193 Bellaterra (Barcelona), Spain\label{aff62}
\and
Institute for Theoretical Particle Physics and Cosmology (TTK), RWTH Aachen University, 52056 Aachen, Germany\label{aff63}
\and
Deutsches Zentrum f\"ur Luft- und Raumfahrt e. V. (DLR), Linder H\"ohe, 51147 K\"oln, Germany\label{aff64}
\and
INAF-Osservatorio Astronomico di Roma, Via Frascati 33, 00078 Monteporzio Catone, Italy\label{aff65}
\and
INFN section of Naples, Via Cinthia 6, 80126, Napoli, Italy\label{aff66}
\and
Institute for Astronomy, University of Hawaii, 2680 Woodlawn Drive, Honolulu, HI 96822, USA\label{aff67}
\and
Dipartimento di Fisica e Astronomia "Augusto Righi" - Alma Mater Studiorum Universit\`a di Bologna, Viale Berti Pichat 6/2, 40127 Bologna, Italy\label{aff68}
\and
Instituto de Astrof\'{\i}sica de Canarias, E-38205 La Laguna, Tenerife, Spain\label{aff69}
\and
European Space Agency/ESRIN, Largo Galileo Galilei 1, 00044 Frascati, Roma, Italy\label{aff70}
\and
Universit\'e Claude Bernard Lyon 1, CNRS/IN2P3, IP2I Lyon, UMR 5822, Villeurbanne, F-69100, France\label{aff71}
\and
UCB Lyon 1, CNRS/IN2P3, IUF, IP2I Lyon, 4 rue Enrico Fermi, 69622 Villeurbanne, France\label{aff72}
\and
Mullard Space Science Laboratory, University College London, Holmbury St Mary, Dorking, Surrey RH5 6NT, UK\label{aff73}
\and
Departamento de F\'isica, Faculdade de Ci\^encias, Universidade de Lisboa, Edif\'icio C8, Campo Grande, PT1749-016 Lisboa, Portugal\label{aff74}
\and
Instituto de Astrof\'isica e Ci\^encias do Espa\c{c}o, Faculdade de Ci\^encias, Universidade de Lisboa, Campo Grande, 1749-016 Lisboa, Portugal\label{aff75}
\and
Department of Astronomy, University of Geneva, ch. d'Ecogia 16, 1290 Versoix, Switzerland\label{aff76}
\and
INFN-Padova, Via Marzolo 8, 35131 Padova, Italy\label{aff77}
\and
Aix-Marseille Universit\'e, CNRS/IN2P3, CPPM, Marseille, France\label{aff78}
\and
INAF-Istituto di Astrofisica e Planetologia Spaziali, via del Fosso del Cavaliere, 100, 00100 Roma, Italy\label{aff79}
\and
Space Science Data Center, Italian Space Agency, via del Politecnico snc, 00133 Roma, Italy\label{aff80}
\and
INFN-Bologna, Via Irnerio 46, 40126 Bologna, Italy\label{aff81}
\and
Institut d'Estudis Espacials de Catalunya (IEEC),  Edifici RDIT, Campus UPC, 08860 Castelldefels, Barcelona, Spain\label{aff82}
\and
Institute of Space Sciences (ICE, CSIC), Campus UAB, Carrer de Can Magrans, s/n, 08193 Barcelona, Spain\label{aff83}
\and
School of Physics, HH Wills Physics Laboratory, University of Bristol, Tyndall Avenue, Bristol, BS8 1TL, UK\label{aff84}
\and
University Observatory, LMU Faculty of Physics, Scheinerstr.~1, 81679 Munich, Germany\label{aff85}
\and
FRACTAL S.L.N.E., calle Tulip\'an 2, Portal 13 1A, 28231, Las Rozas de Madrid, Spain\label{aff86}
\and
INFN-Sezione di Milano, Via Celoria 16, 20133 Milano, Italy\label{aff87}
\and
Institute of Theoretical Astrophysics, University of Oslo, P.O. Box 1029 Blindern, 0315 Oslo, Norway\label{aff88}
\and
Felix Hormuth Engineering, Goethestr. 17, 69181 Leimen, Germany\label{aff89}
\and
Technical University of Denmark, Elektrovej 327, 2800 Kgs. Lyngby, Denmark\label{aff90}
\and
Cosmic Dawn Center (DAWN), Denmark\label{aff91}
\and
Max-Planck-Institut f\"ur Astronomie, K\"onigstuhl 17, 69117 Heidelberg, Germany\label{aff92}
\and
NASA Goddard Space Flight Center, Greenbelt, MD 20771, USA\label{aff93}
\and
Department of Physics and Astronomy, University College London, Gower Street, London WC1E 6BT, UK\label{aff94}
\and
Department of Physics and Helsinki Institute of Physics, Gustaf H\"allstr\"omin katu 2, University of Helsinki, 00014 Helsinki, Finland\label{aff95}
\and
Department of Physics, P.O. Box 64, University of Helsinki, 00014 Helsinki, Finland\label{aff96}
\and
Helsinki Institute of Physics, Gustaf H{\"a}llstr{\"o}min katu 2, University of Helsinki, 00014 Helsinki, Finland\label{aff97}
\and
Laboratoire d'etude de l'Univers et des phenomenes eXtremes, Observatoire de Paris, Universit\'e PSL, Sorbonne Universit\'e, CNRS, 92190 Meudon, France\label{aff98}
\and
SKAO, Jodrell Bank, Lower Withington, Macclesfield SK11 9FT, UK\label{aff99}
\and
Centre de Calcul de l'IN2P3/CNRS, 21 avenue Pierre de Coubertin 69627 Villeurbanne Cedex, France\label{aff100}
\and
Universit\"at Bonn, Argelander-Institut f\"ur Astronomie, Auf dem H\"ugel 71, 53121 Bonn, Germany\label{aff101}
\and
INFN-Sezione di Roma, Piazzale Aldo Moro, 2 - c/o Dipartimento di Fisica, Edificio G. Marconi, 00185 Roma, Italy\label{aff102}
\and
Department of Physics, Institute for Computational Cosmology, Durham University, South Road, Durham, DH1 3LE, UK\label{aff103}
\and
Universit\'e Paris Cit\'e, CNRS, Astroparticule et Cosmologie, 75013 Paris, France\label{aff104}
\and
CNRS-UCB International Research Laboratory, Centre Pierre Bin\'etruy, IRL2007, CPB-IN2P3, Berkeley, USA\label{aff105}
\and
Institut d'Astrophysique de Paris, 98bis Boulevard Arago, 75014, Paris, France\label{aff106}
\and
Telespazio UK S.L. for European Space Agency (ESA), Camino bajo del Castillo, s/n, Urbanizacion Villafranca del Castillo, Villanueva de la Ca\~nada, 28692 Madrid, Spain\label{aff107}
\and
Institut de F\'{i}sica d'Altes Energies (IFAE), The Barcelona Institute of Science and Technology, Campus UAB, 08193 Bellaterra (Barcelona), Spain\label{aff108}
\and
School of Mathematics and Physics, University of Surrey, Guildford, Surrey, GU2 7XH, UK\label{aff109}
\and
European Space Agency/ESTEC, Keplerlaan 1, 2201 AZ Noordwijk, The Netherlands\label{aff110}
\and
DARK, Niels Bohr Institute, University of Copenhagen, Jagtvej 155, 2200 Copenhagen, Denmark\label{aff111}
\and
Waterloo Centre for Astrophysics, University of Waterloo, Waterloo, Ontario N2L 3G1, Canada\label{aff112}
\and
Department of Physics and Astronomy, University of Waterloo, Waterloo, Ontario N2L 3G1, Canada\label{aff113}
\and
Perimeter Institute for Theoretical Physics, Waterloo, Ontario N2L 2Y5, Canada\label{aff114}
\and
Universit\'e Paris-Saclay, Universit\'e Paris Cit\'e, CEA, CNRS, AIM, 91191, Gif-sur-Yvette, France\label{aff115}
\and
Centre National d'Etudes Spatiales -- Centre spatial de Toulouse, 18 avenue Edouard Belin, 31401 Toulouse Cedex 9, France\label{aff116}
\and
Dipartimento di Fisica e Astronomia "G. Galilei", Universit\`a di Padova, Via Marzolo 8, 35131 Padova, Italy\label{aff117}
\and
Institut f\"ur Theoretische Physik, University of Heidelberg, Philosophenweg 16, 69120 Heidelberg, Germany\label{aff118}
\and
Institut de Recherche en Astrophysique et Plan\'etologie (IRAP), Universit\'e de Toulouse, CNRS, UPS, CNES, 14 Av. Edouard Belin, 31400 Toulouse, France\label{aff119}
\and
Universit\'e St Joseph; Faculty of Sciences, Beirut, Lebanon\label{aff120}
\and
Departamento de F\'isica, FCFM, Universidad de Chile, Blanco Encalada 2008, Santiago, Chile\label{aff121}
\and
Universit\"at Innsbruck, Institut f\"ur Astro- und Teilchenphysik, Technikerstr. 25/8, 6020 Innsbruck, Austria\label{aff122}
\and
Satlantis, University Science Park, Sede Bld 48940, Leioa-Bilbao, Spain\label{aff123}
\and
Infrared Processing and Analysis Center, California Institute of Technology, Pasadena, CA 91125, USA\label{aff124}
\and
Instituto de Astrof\'isica e Ci\^encias do Espa\c{c}o, Faculdade de Ci\^encias, Universidade de Lisboa, Tapada da Ajuda, 1349-018 Lisboa, Portugal\label{aff125}
\and
Cosmic Dawn Center (DAWN)\label{aff126}
\and
Niels Bohr Institute, University of Copenhagen, Jagtvej 128, 2200 Copenhagen, Denmark\label{aff127}
\and
Universidad Polit\'ecnica de Cartagena, Departamento de Electr\'onica y Tecnolog\'ia de Computadoras,  Plaza del Hospital 1, 30202 Cartagena, Spain\label{aff128}
\and
Dipartimento di Fisica e Scienze della Terra, Universit\`a degli Studi di Ferrara, Via Giuseppe Saragat 1, 44122 Ferrara, Italy\label{aff129}
\and
Istituto Nazionale di Fisica Nucleare, Sezione di Ferrara, Via Giuseppe Saragat 1, 44122 Ferrara, Italy\label{aff130}
\and
INAF, Istituto di Radioastronomia, Via Piero Gobetti 101, 40129 Bologna, Italy\label{aff131}
\and
Universit\'e C\^{o}te d'Azur, Observatoire de la C\^{o}te d'Azur, CNRS, Laboratoire Lagrange, Bd de l'Observatoire, CS 34229, 06304 Nice cedex 4, France\label{aff132}
\and
ICSC - Centro Nazionale di Ricerca in High Performance Computing, Big Data e Quantum Computing, Via Magnanelli 2, Bologna, Italy\label{aff133}
\and
SCITAS, Ecole Polytechnique F\'ed\'erale de Lausanne (EPFL), 1015 Lausanne, Switzerland\label{aff134}
\and
Instituto de F\'isica Te\'orica UAM-CSIC, Campus de Cantoblanco, 28049 Madrid, Spain\label{aff135}
\and
CEA Saclay, DFR/IRFU, Service d'Astrophysique, Bat. 709, 91191 Gif-sur-Yvette, France\label{aff136}
\and
Universit\'e PSL, Observatoire de Paris, Sorbonne Universit\'e, CNRS, LERMA, 75014, Paris, France\label{aff137}
\and
Universit\'e Paris-Cit\'e, 5 Rue Thomas Mann, 75013, Paris, France\label{aff138}
\and
Univ. Grenoble Alpes, CNRS, Grenoble INP, LPSC-IN2P3, 53, Avenue des Martyrs, 38000, Grenoble, France\label{aff139}
\and
Dipartimento di Fisica, Sapienza Universit\`a di Roma, Piazzale Aldo Moro 2, 00185 Roma, Italy\label{aff140}
\and
Aurora Technology for European Space Agency (ESA), Camino bajo del Castillo, s/n, Urbanizacion Villafranca del Castillo, Villanueva de la Ca\~nada, 28692 Madrid, Spain\label{aff141}
\and
Zentrum f\"ur Astronomie, Universit\"at Heidelberg, Philosophenweg 12, 69120 Heidelberg, Germany\label{aff142}
\and
Dipartimento di Fisica - Sezione di Astronomia, Universit\`a di Trieste, Via Tiepolo 11, 34131 Trieste, Italy\label{aff143}
\and
ICL, Junia, Universit\'e Catholique de Lille, LITL, 59000 Lille, France\label{aff144}
\and
CERCA/ISO, Department of Physics, Case Western Reserve University, 10900 Euclid Avenue, Cleveland, OH 44106, USA\label{aff145}
\and
Laboratoire Univers et Th\'eorie, Observatoire de Paris, Universit\'e PSL, Universit\'e Paris Cit\'e, CNRS, 92190 Meudon, France\label{aff146}
\and
Departamento de F{\'\i}sica Fundamental. Universidad de Salamanca. Plaza de la Merced s/n. 37008 Salamanca, Spain\label{aff147}
\and
Universit\'e de Strasbourg, CNRS, Observatoire astronomique de Strasbourg, UMR 7550, 67000 Strasbourg, France\label{aff148}
\and
Center for Data-Driven Discovery, Kavli IPMU (WPI), UTIAS, The University of Tokyo, Kashiwa, Chiba 277-8583, Japan\label{aff149}
\and
California Institute of Technology, 1200 E California Blvd, Pasadena, CA 91125, USA\label{aff150}
\and
Department of Mathematics and Physics E. De Giorgi, University of Salento, Via per Arnesano, CP-I93, 73100, Lecce, Italy\label{aff151}
\and
INFN, Sezione di Lecce, Via per Arnesano, CP-193, 73100, Lecce, Italy\label{aff152}
\and
INAF-Sezione di Lecce, c/o Dipartimento Matematica e Fisica, Via per Arnesano, 73100, Lecce, Italy\label{aff153}
\and
Kapteyn Astronomical Institute, University of Groningen, PO Box 800, 9700 AV Groningen, The Netherlands\label{aff154}
\and
Instituto de F\'isica de Cantabria, Edificio Juan Jord\'a, Avenida de los Castros, 39005 Santander, Spain\label{aff155}
\and
Department of Computer Science, Aalto University, PO Box 15400, Espoo, FI-00 076, Finland\label{aff156}
\and
 Instituto de Astrof\'{\i}sica de Canarias, E-38205 La Laguna; Universidad de La Laguna, Dpto. Astrof\'\i sica, E-38206 La Laguna, Tenerife, Spain\label{aff157}
\and
Ruhr University Bochum, Faculty of Physics and Astronomy, Astronomical Institute (AIRUB), German Centre for Cosmological Lensing (GCCL), 44780 Bochum, Germany\label{aff158}
\and
Department of Physics and Astronomy, Vesilinnantie 5, University of Turku, 20014 Turku, Finland\label{aff159}
\and
Finnish Centre for Astronomy with ESO (FINCA), Quantum, Vesilinnantie 5, University of Turku, 20014 Turku, Finland\label{aff160}
\and
Serco for European Space Agency (ESA), Camino bajo del Castillo, s/n, Urbanizacion Villafranca del Castillo, Villanueva de la Ca\~nada, 28692 Madrid, Spain\label{aff161}
\and
ARC Centre of Excellence for Dark Matter Particle Physics, Melbourne, Australia\label{aff162}
\and
Centre for Astrophysics \& Supercomputing, Swinburne University of Technology,  Hawthorn, Victoria 3122, Australia\label{aff163}
\and
Department of Physics and Astronomy, University of the Western Cape, Bellville, Cape Town, 7535, South Africa\label{aff164}
\and
DAMTP, Centre for Mathematical Sciences, Wilberforce Road, Cambridge CB3 0WA, UK\label{aff165}
\and
Kavli Institute for Cosmology Cambridge, Madingley Road, Cambridge, CB3 0HA, UK\label{aff166}
\and
Department of Astrophysics, University of Zurich, Winterthurerstrasse 190, 8057 Zurich, Switzerland\label{aff167}
\and
Department of Physics, Centre for Extragalactic Astronomy, Durham University, South Road, Durham, DH1 3LE, UK\label{aff168}
\and
IRFU, CEA, Universit\'e Paris-Saclay 91191 Gif-sur-Yvette Cedex, France\label{aff169}
\and
Oskar Klein Centre for Cosmoparticle Physics, Department of Physics, Stockholm University, Stockholm, SE-106 91, Sweden\label{aff170}
\and
Astrophysics Group, Blackett Laboratory, Imperial College London, London SW7 2AZ, UK\label{aff171}
\and
Centro de Astrof\'{\i}sica da Universidade do Porto, Rua das Estrelas, 4150-762 Porto, Portugal\label{aff172}
\and
Instituto de Astrof\'isica e Ci\^encias do Espa\c{c}o, Universidade do Porto, CAUP, Rua das Estrelas, PT4150-762 Porto, Portugal\label{aff173}
\and
HE Space for European Space Agency (ESA), Camino bajo del Castillo, s/n, Urbanizacion Villafranca del Castillo, Villanueva de la Ca\~nada, 28692 Madrid, Spain\label{aff174}
\and
INAF - Osservatorio Astronomico d'Abruzzo, Via Maggini, 64100, Teramo, Italy\label{aff175}
\and
Theoretical astrophysics, Department of Physics and Astronomy, Uppsala University, Box 516, 751 37 Uppsala, Sweden\label{aff176}
\and
Mathematical Institute, University of Leiden, Einsteinweg 55, 2333 CA Leiden, The Netherlands\label{aff177}
\and
Institute of Astronomy, University of Cambridge, Madingley Road, Cambridge CB3 0HA, UK\label{aff178}
\and
Univ. Lille, CNRS, Centrale Lille, UMR 9189 CRIStAL, 59000 Lille, France\label{aff179}
\and
Institute for Particle Physics and Astrophysics, Dept. of Physics, ETH Zurich, Wolfgang-Pauli-Strasse 27, 8093 Zurich, Switzerland\label{aff180}
\and
Department of Astrophysical Sciences, Peyton Hall, Princeton University, Princeton, NJ 08544, USA\label{aff181}
\and
Space physics and astronomy research unit, University of Oulu, Pentti Kaiteran katu 1, FI-90014 Oulu, Finland\label{aff182}
\and
Department of Physics and Astronomy, Lehman College of the CUNY, Bronx, NY 10468, USA\label{aff183}
\and
American Museum of Natural History, Department of Astrophysics, New York, NY 10024, USA\label{aff184}
\and
International Centre for Theoretical Physics (ICTP), Strada Costiera 11, 34151 Trieste, Italy\label{aff185}
\and
Center for Computational Astrophysics, Flatiron Institute, 162 5th Avenue, 10010, New York, NY, USA\label{aff186}}    

%
%
 \abstract{We present 72 additional galaxy-galaxy strong lenses that complement the sample discovered in the Euclid Quick Release 1 data ($63.1\,\deg^2$) of the Strong Lens Discovery Engine (SLDE) papers A–E. It is shown that previous pre-selection of potential lenses, which excluded objects from the \textit{Gaia} catalogue, led to missing several bright and low-redshift strong lenses, adding more than $10\%$ new strong lens candidates compared to the previous search. In total, the catalogue includes 38 ``grade A'' (confident) and 34 ``grade B'' (probable) candidates. These lenses are identified through a combination of two independent searches for bright nearby objects: one based on machine-learning models followed by expert visual inspection, and the other based solely on expert visual inspection, targeting objects not included in the initial machine-learning selection (a limitation identified only after extensive visual inspection). With these additional strong lens candidates, we augment the expected number of high-confidence candidates in the \textit{Euclid} Wide Survey from previous forecasts to \num{120000}. Detailed semi-automated lens modelling confirms at least 41 systems out of 72, a fraction consistent with that found in SLDE A (315 out of 488). These include: multiple edge-on disc lenses; sources with arcs near the lens centre; ``red sources''; and an edge-on disk galaxy lensing a galaxy merger, producing two sets of lensed features, an Einstein ring and a doubly imaged component. The median redshift of these systems is $\Delta z \sim 0.3$ lower than that of the SLDE A sample.}
%
%
%
   \titlerunning{\Euclid\/ Q1: The Strong Lensing Discovery Engine F}
   \authorrunning{Euclid Collaboration: L. R. Ecker et al.}
   
   \maketitle
%
%
%
%
   
\section{\label{sc:Intro}Introduction}
The ESA \Euclid mission is set to image approximately $\num{14000}\, \deg^2$ of the extragalactic sky and detect over 1.5 billion galaxies (\citealp{EuclidSkyOverview}). Among these, forecasts predict the discovery of approximately $\num{100000}$ galaxy–galaxy strong lens systems with space-based imaging (\citealp{Collett_2015}), which is about two orders of magnitude more than currently known from a single survey. Strong lenses provide a unique window into a wide range of astrophysical and cosmological questions; they enable precise measurements of the Hubble constant $H_0$ and tests of cosmological parameters and dark energy (e.g. \citealt{Refsdal_1964, Suyu_2013, suyu2017h0licow, suyu2020holismokes, Queirolo_23}), constraints on the total mass and dark matter distribution of galaxies (e.g. \citealt{Koopmans_2006, Auger_2010}), detailed studies of dark matter distribution and substructure (e.g. \citealt{Vegetti_2010, Despali_2021, Minor_2021, oriordan2024angularcomplexitystronglens}), and the magnified observation of high-redshift galaxies, now extending to $z \gtrsim 10$ with JWST (e.g. \citealt{Furtak_2022, Atek_2023}).

Thanks to its high spatial resolution and stable space-based point spread-function (PSF), \Euclid is expected to resolve lenses with Einstein radii as small as $\ang{;;0.6}$--$\ang{;;0.7}$ -- the peak of the predicted distribution (\citealp{Collett_2015}, \citealp{Sonnenfeld_2023}) -- outperforming ground-based surveys in sensitivity to compact lens configurations. While many lenses will be well characterised using Euclid imaging alone, certain science cases will require complementary or higher-resolution follow-up observations, for example with HST, JWST, Roman, or CSST, to provide additional wavelength coverage, multi-band constraints, or time-domain information for variable sources.

\Euclid{}’s wide coverage will also enable systematic studies of rare lensing configurations across diverse categories, including double-source-plane lenses \citep[DSPLs;][]{Q1-SP054}, lensed quasars, low-redshift systems, and edge-on disc lenses. Previous surveys with lens searches such as DES (e.g. \citealp{Rojas_2022, O_Donnell_2022}), KiDS (e.g. \citealp{2018MNRAS.480.1163S, Petrillo_2017, Petrillo_2019, Grespan_2024, nagam2025, Li_2020_Kids, Li_2021_Kids}), and PanSTARRS (\citealp{Ca_ameras_2020}) detected around $\sim$0.1 lenses per $\deg^2$, while deeper imaging from lens searches with Hyper Suprime-Cam (HSC; e.g. \citealp{Sonnenfeld_2020_HSC, Ca_ameras_2021, Shu_2022, Schuldt_2025, Jaelani_2024}) reached about 1 lens per $\deg^2$. Preliminary visual searches of the \Euclid Early Release Observations (ERO) have already revealed up to 10 lenses per $\deg^2$, showcasing the mission’s transformative potential (e.g. \citealp{Pearce-Casey24, Nagam25, AcevedoBarroso24}). To systematically identify these systems, the Strong Lens Discovery Engine (SLDE) was developed as a dedicated framework for detecting strong lenses in \Euclid{} data.

This paper extends the series of the Strong Lens Discovery Engine (SLDE), supplementing the efforts to identify strong lenses in the Euclid Quick Release 1 (Q1). SLDE A (\citealp{Q1-SP048}) presented the initial SLDE lens search and catalogue, followed by SLDE B–E (\citealp{Q1-SP052, Q1-SP053, Q1-SP054, Q1-SP059}). Due to the selection applied in these previous papers, a particular subset of very bright and lower in redshifts galaxies remained largely unexplored. This paper complements the earlier searches by targeting on this class of lenses. 

This paper is structured as follows. In Sect.~\ref{sc:Data}, we discuss the \Euclid Q1 data products used in this work, as well as those employed for visual inspection. The methodology for identifying strong lenses is presented in Sect.~\ref{sc:Method}. In here, we describe the \texttt{Zoobot} selection process, which was partially followed by human visual inspection. Section~\ref{sec:Modelling} describes the methodology to model these lenses and individual objects are discussed. Section~\ref{sec:Redshifts} presents a selection of spectra of the lens galaxies. We summarise the overall workflow and discuss systematic biases relative to SLDE A in Sect.~\ref{sec:Results}. Finally, the conclusions are given in Sect.~\ref{sec:Conclusion}. When required, we assume a flat $\Lambda$CDM cosmology with $\Omega_{\rm m} = 0.311$ and $H_0 = 67.66\,\mathrm{km\,s^{-1}\,Mpc^{-1}}$, consistent with \citet{Planck_2018}.

\section{\label{sc:Data}Data}
We utilized imaging from the \Euclid VIS and NISP instruments \citep{EuclidSkyVIS, EuclidSkyNISP}, as processed for the Q1 data release \citep{Q1cite, Q1-TP002, Q1-TP003, Q1-TP004}. VIS delivers broad-band optical imaging over approximately 550--900~nm, with a pixel scale of  $\ang{;;0.1}\,\rm{pixel}^{-1}$ and a field of view of $0.57\,\deg^2$ \citep{EuclidSkyVIS}. The Q1 release includes reduced VIS images, detection and photometry catalogues, segmentation maps, a position-dependent PSF, and higher-level products such as photometric redshifts and morphological classifications. Complementing VIS, the NISP instrument provides near-infrared imaging in the 950--2020 nm range \citep{EuclidSkyNISP}.

The data were accessed via the ESA Science Archive Service and the ESA Datalabs platform\footnote{\url{https://www.cosmos.esa.int/web/euclid/data-access}}. For each candidate, a $\ang{;;10} \times \ang{;;10}$ cutout (100$\times$100 MER mosaicked pixels at $\ang{;;0.1}\,\rm{pixel}^{-1}$) was extracted. Hereafter VIS and the three NISP bands are referred to as \IE, \YE, \JE, and \HE.

\section{Method}
\label{sc:Method}

\subsection{Cutout preparation}
\begin{figure}[htbp!]
\centering
\includegraphics[angle=0,width=1.0\hsize]{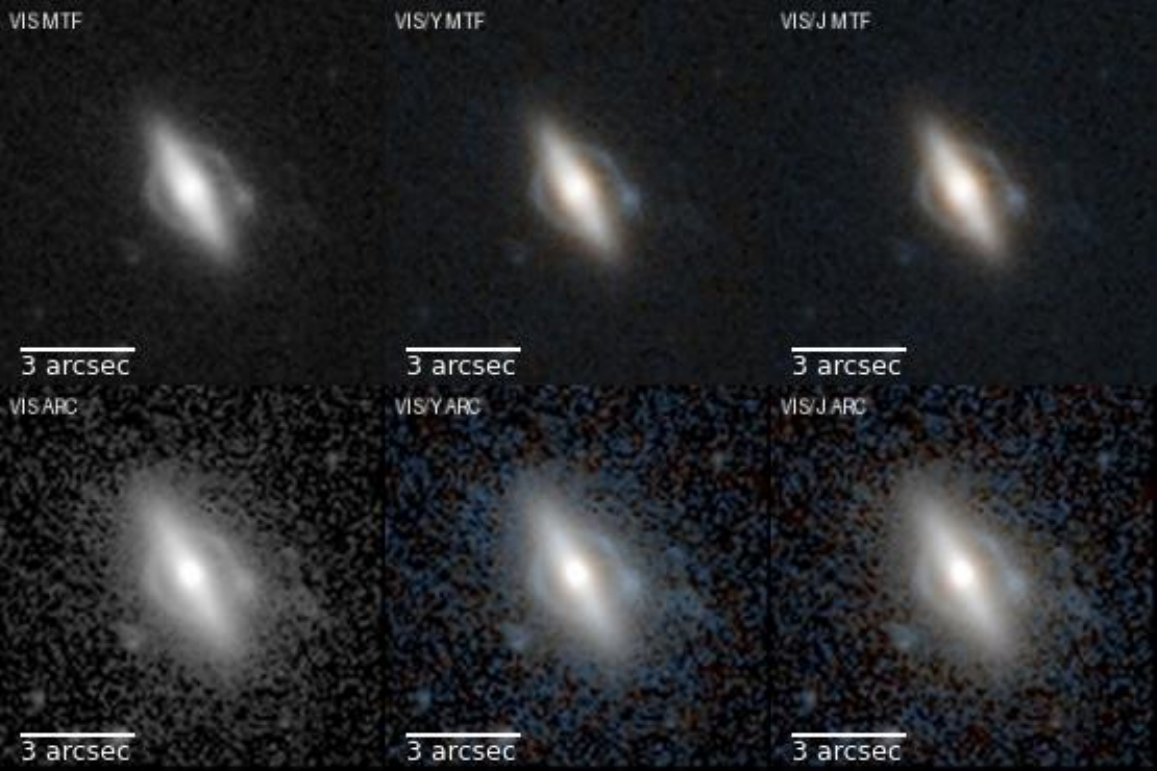}
\caption{Illustration of image processing configurations. Each panel shows a different combination of intensity scaling (arcsinh or midtone transfer function) and colour mapping (greyscale \IE; \YE/median/\IE; \JE/median/\IE). In all cases, luminance is defined using the \IE image to maximise resolution.
}
\label{fig:ScalingImages}
\end{figure}

Due to storage constraints associated with large-scale surveys, a practical solution commonly used in earlier work of the SLDE series is to generate compressed JPG images from the original FITS files. To ensure that visual inspectors can effectively identify morphological structures and potential strong lenses, different visualisation schemes were used (Fig. \ref{fig:ScalingImages}). For the ML algorithm, we used greyscale images in the \IE band, with raw flux values rescaled using an arcsinh transformation,
\begin{equation}
x^\prime = \sinh^{-1}(Qx) \; ,
\end{equation}
where $x$ is the raw flux and $Q$ controls the strength of the transformation (\citealp{Lupton2004}). This scaling enhances faint features while preserving bright structures, making the images better suited for the algorithm. For the human visual inspection we adopted $Q = 500/1/0.5$ for the \IE/\YE/\JE\ bands (as in SLDE A), respectively. Additionally, the human visual inspection included a midtone transfer function (MTF) scaling, and additional band combinations (\IE \& \YE and \IE \& \JE) to enhance feature visibility,
\begin{equation}
x^\prime = \frac{(m-1)x}{(2m - 1)x - m} \; ,
\end{equation}
where the stretch factor $m$ was automatically chosen such that each cutout had a mean scaled value of 0.2. In both methods, we applied a 99.85th percentile clip to suppress outlier flux values before rescaling. Figure \ref{fig:ScalingImages} shows a representative example of the image variants available to the visual inspectors. Additionally, \HE-band images were used to verify the best fitting lensing model obtained through the \IE band.\\

\subsection{Initial visual inspection of extended bright galaxies} 
\label{sec: fabricius_selection}
\begin{figure}[htbp!]
\centering
\includegraphics[angle=0,width=1.0\hsize]{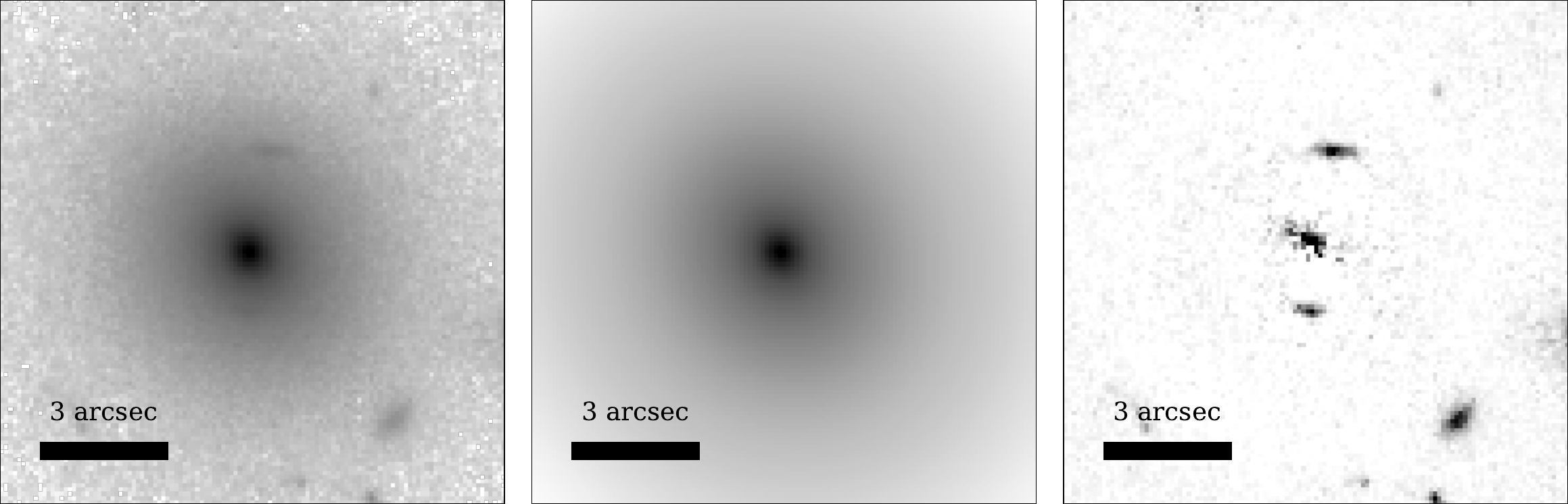}
\caption{Example showcasing the benefits of subtracting the light of the foreground galaxy. \emph{Left}: The original image of a strong lens candidate in grey scale in the VIS filter. \emph{Centre}: The model of the foreground lens light with multi-Gaussian Expansion. \emph{Right}: Residual image obtained by subtracting the model image from the original image. Two bright arcs are visible.}
\label{fig:1}
\end{figure}

Several strong lens candidates in Q1 were identified in the double nuclei search (\citealp{Q1-SP066}) that were not found by SLDE A. A key systematic in the previous SLDE A selection was the exclusion of all objects with a {\tt gaia\_id}, which removed not only foreground stars but also bright, nearby galaxies detectable by \textit{Gaia}, leading to missed lens candidates. In this work, we instead cross-matched sources only with the \textit{Gaia} star catalogue (\citealp{Gaia2023}) and discarded confirmed stars. The selection for this inspection applied the thresholds
\begin{itemize}
\item {\tt flux\_vis\_1fwhm\_aper} $>$ 20 (corresponding to \IE $<$ 20.6);
\item {\tt segmentation\_area} $>$ 5500.
\end{itemize}
Sources meeting these criteria and not flagged as stars resulted in $\num{15180}$ objects. Initial modelling and light subtraction produced residual images, which visual inspection indicated contained structures potentially indicative of strong lensing in 160 systems. Fig.~\ref{fig:1} shows an example of lens light subtraction for one of these candidates. We also reran the \texttt{Zoobot} classifier \citep{Q1-SP053, Walmsley2023} (from SLDE A) on objects with \textit{Gaia} IDs and recovered high-ranking lens candidates, confirming that the previous omission was due to the {\tt gaia\_id} selection cut rather than a bias against faint counter-arcs.

\subsection{Extension to selection cuts from SLDE A}
To investigate the impact of the \textit{Gaia} exclusion, we later repeated the selection process by including objects with a {\tt gaia\_id} from the SLDE A. We adopted the same selection criteria as in the SLDE A search (\citealp{Q1-SP048}) to construct our strong lens candidate sample from the Q1 MER catalogue (\citealp{Q1-TP004}). So the selection criteria for this work (by including the {\tt GAIA\_CROSSMATCH}) are:
\begin{itemize}
    \item {\tt FLUX\_DETECTION\_TOTAL} $\geq$ 3.63078 (corresponding to \IE < 22.5);
    \item {\tt SEGMENTATION\_AREA} $\geq$ 259 pixels;
    \item {\tt VIS\_DET} = 1;
    \item {\tt MUMAX\_MINUS\_MAG} $\geq$ $-2.6$;
    \item {\tt MU\_MAX} $\geq$ 15.0;
    \item {\tt SPURIOUS\_PROB} $<$ 0.05;
    \item {\tt GAIA\_CROSSMATCH}.
\end{itemize}
Using this second set of selection criteria, we identified an additional \num{57536} objects that were missed by SLDE A. These were cross-matched with the DR3 \textit{Gaia} star catalogue (\citealp{Gaia2023}). The initial selection (from the double nucleii project) had yielded $\num{15180}$ candidates, and there was significant overlap between the two samples. After removing previously inspected objects from the new selection and stars from the DR3 \textit{Gaia} star catalogue, we obtained an additional $\num{12787}$ unique images. 

Since the initial sample of $\num{15180}$ objects had already undergone visual inspection, it was treated as complete. For the remaining $\num{12787}$ newly identified objects, we aimed to reduce the number requiring expert visual inspection by applying two complementary preselection strategies: an ML classifier and a citizen-science search using Space Warps \citep{Marshall_zooniverse, More_zooniverse}, a project designed to identify strong lenses \citep{geach2015red, Sonnenfeld_2020_HSC, Gonzalez2025}. Specifically, the ML-based ranking used the best-performing network identified in \citet{Q1-SP053}, a fine-tuned version of \texttt{Zoobot}, to assign lensing probabilities (Sect.~\ref{sec:ML}). The highest-scoring candidates from both the ML analysis and the citizen-science inspection were then forwarded to expert visual inspection, ensuring a consistent and homogeneous evaluation across all objects.

\subsection{Candidate preselection overview}
\label{sec:ML}

Our approach follows the strategy outlined in \citet{Pearce-Casey24}, who demonstrated the importance of using realistic training data and foundation models for identifying strong lenses in \Euclid data. They also showed that the \texttt{Zoobot} model \citep{Walmsley2023}, pre-trained on galaxy morphology tasks using real survey data, achieved the best performance among all submissions, even without fine-tuning. This result was confirmed again in \citet{Q1-SP053}, while \citet{Q1-SP048} provided a catalogue of 497 strong lens candidates. As discussed in Sect.~\ref{sec: fabricius_selection}, this catalogue exhibits a systematic bias against bright and low-redshift galaxies. To ensure that the retrained \texttt{Zoobot} model used here does not inherit that bias, we supplemented the training set with a random selection of high-confidence candidates from the initial list of 160 objects. This updated training set enhanced the model's ability to generalize to a wider range of morphologies, as first demonstrated in \citet{Q1-SP085} and also shown in Xu et al. (in prep).

\begin{figure}[htbp!]
\centering
\includegraphics[angle=0,width=1.0\hsize]{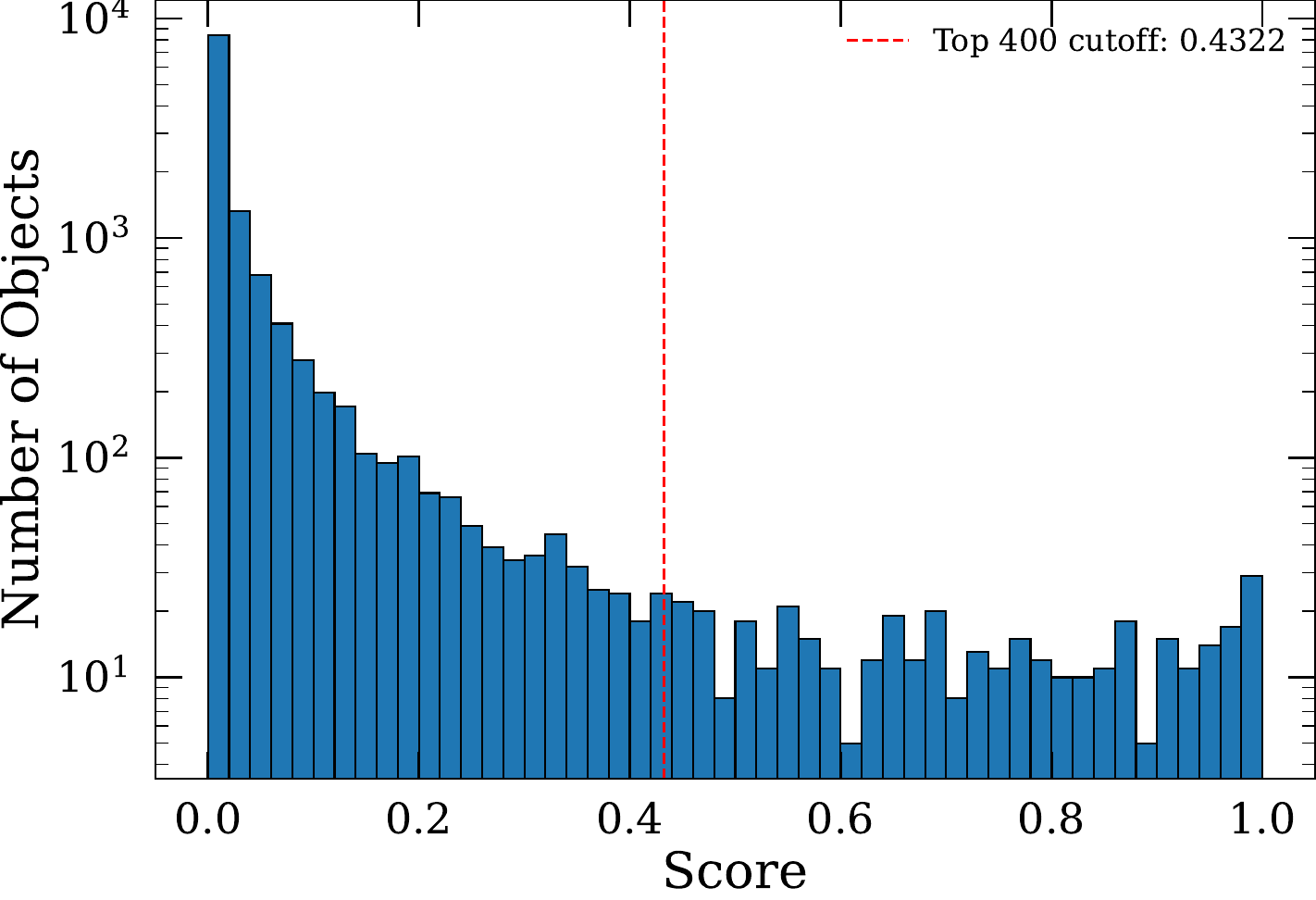}
\caption{Distribution of scores from \texttt{Zoobot}. The vertical dashed line indicates the cutoff chosen where all objects right from this line were chosen to be visually inspected.}
\label{fig:Zoobot_Distribution}
\end{figure}

For the present study, candidate selection via ML was based entirely on the predictions of this retrained \texttt{Zoobot} model. From the resulting ranked list, we selected the top 400 highest-scoring candidates, corresponding to a minimum score of 0.432. These were then injected into the Zooniverse platform \citep{Marshall_zooniverse, More_zooniverse}, where they were subsequently evaluated by expert astronomers. The score distribution and the selected cutoff are illustrated in Fig.~\ref{fig:Zoobot_Distribution}.

In parallel to the ML-based selection, we also conducted an independent candidate search based on citizen-science visual inspection. This effort yielded an additional sample of 96 strong lens candidates, which were subsequently subjected to the same expert visual inspection and grading procedure described in Sect.~\ref{sec:VIExpert}. In parallel, an independent citizen-science visual inspection was carried out on a separate set of candidates originating from this work, as well as on an entirely independent candidate list produced by Vincken et al. (in prep.). While the two candidate samples are disjoint, both were inspected using the same citizen-science framework, enabling a consistent comparison between volunteer and expert classifications (Holloway et al., in prep.). to directly compare the classifications of volunteer citizen scientists with those of expert strong-lens graders.

\subsection{\label{sec:VIExpert}Expert visual inspection overview}
Following ML preselection, we conducted a visual inspection with experts on a mosaic of 2$\times$3 images with \IE-only, \IE+\YE, and \IE+\JE colour images in MTF and arcsinh scaling. Figure \ref{fig:ScalingImages} shows an example of this. These were presented in the Galaxy Judges platform, where each image was graded independently by 19 experts. Our expert annotations were provided by professional astronomers from the Euclid Consortium’s Strong Lensing Science Working Group and the Max-Planck Institute for extraterrestrial physics (MPE). In total, 28 volunteers contributed \num{6550} annotations across \num{655} images. Each expert was randomly assigned images from the candidate pool, and images were considered `retired' once they had received ten independent annotations.

\begin{figure*}[h!]
\centering
\includegraphics[width=\textwidth]{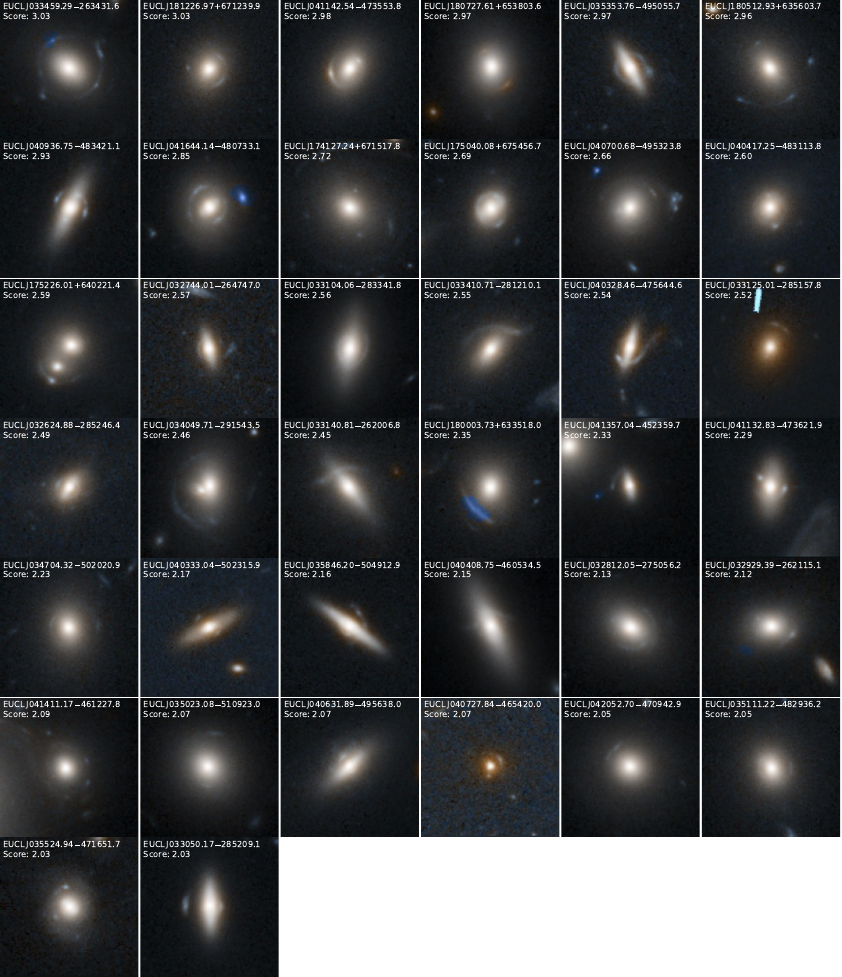}
\caption{Mosaic of strong lens candidates with grade A. The images were created with an MTF scaling of \IE+\JE. }
\label{fig:grade_A}
\end{figure*}

\begin{figure*}[!]
\centering
\includegraphics[width=\textwidth]{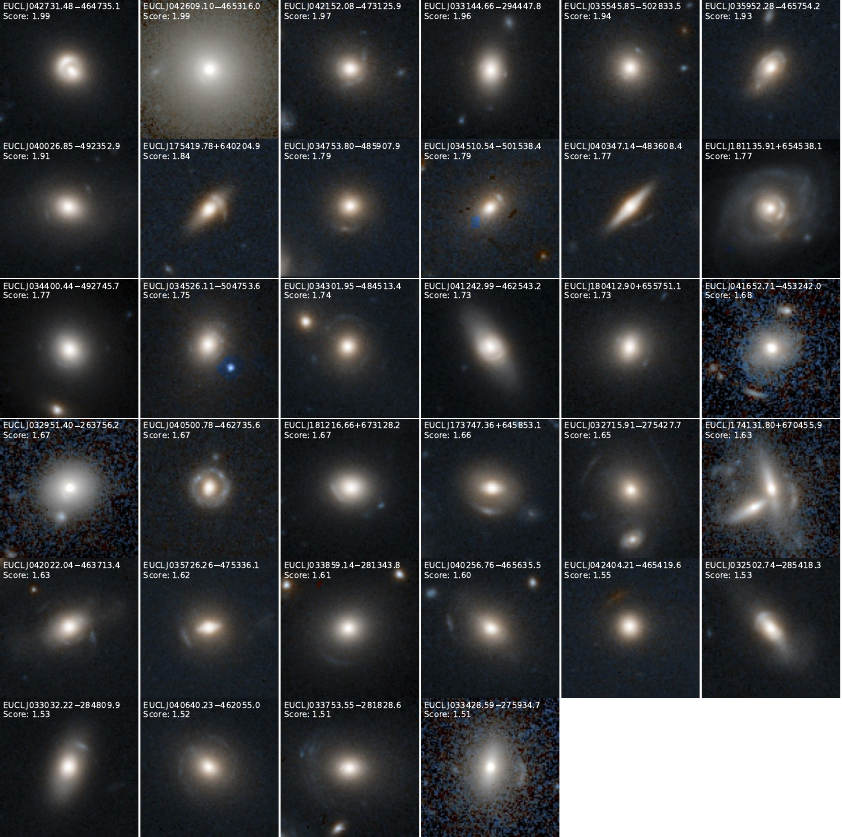}
\caption{As in Fig.~\ref{fig:grade_A} but for strong lens candidates with grade B. }
\label{fig:grade_B}
\end{figure*}

This followed a similar procedure as the visual inspection from \citet{Q1-SP048}, so no simulation set was needed. While simulations can be used to validate whether experts can reliably recognize and recover lens systems (e.g. \citealp{AcevedoBarroso24}), the graders in this work were already experienced from SLDE A, making such a validation unnecessary. Each expert graded the objects according to the scoring system described in \citet{Ca_ameras_2020}, following the procedure outlined in \citet{AcevedoBarroso24}
\begin{itemize}
  \item  Grade A+ score value 3;\\
  \textit{Confident lens:} particular scientific value;
  \item Grade A score value 3;\\ \textit{Confident lens}: clear lensing features are present; no further information is required;  
  \item Grade B score value 2;\\ \textit{Probable lens}: lensing features are visible, but additional data are needed for confirmation;  
  \item Grade C score value 1;\\ \textit{Possible lens}: features may suggest lensing but can also be explained otherwise; 
  \item X score value 0;\\ \textit{Not a lens.}
\end{itemize}
We assign a final score to each lens by first applying a correction to account for systematic optimism or bias in each expert’s scoring, and then averaging the reweighted scores from the ten experts who evaluated each image (see SLDE A). Grade A and grade B lenses, together with their visual inspection scores, are shown in Fig.~\ref{fig:grade_A} and Fig.~\ref{fig:grade_B}, respectively. After visual inspection, candidates with an average grade above 1.5 (corresponding to grade A and grade B lenses) were forwarded to a dedicated lens modelling pipeline for physical validation. The grade A candidates are displayed in Fig.~\ref{fig:grade_A}, while the grade B candidates are displayed in Fig.~\ref{fig:grade_B}.

\subsection{Search results}
Below, we summarise the number of objects from the lens searches.
\begin{enumerate}
    \item Sources in Q1: \num{29767644}
    \begin{enumerate}
        \item Candidates passing selection cuts from \citep{Q1-SP066}: \num{15180}
        \item Candidates marked as candidates by MF, SS, and LRE: 160
        \item Candidates passing selection cuts from \citet{Q1-SP048} with {\tt gaia\_id} outside of \citep{Q1-SP066}: \num{12787}
        \item Candidates searched by ML: \num{12593}
        \item Candidates searched by Citizens: \num{12503}
    \end{enumerate}
    \item Candidates shown to experts
    \begin{enumerate}
        \item ML only: \num{399}
        \item Citizens: \num{96}
        \item Candidates from \citep{Q1-SP066}: \num{160}
        \item Total: 655
    \end{enumerate}
    \item Expert grades
    \begin{enumerate}
        \item Grade A: 38
        \item Grade B: 34
    \end{enumerate}
    \item Modelling of grade A and B candidates
    \begin{enumerate} 
        \item Successful fit, likely lens: 41
        \item Successful fit, unlikely lens: 11
        \item Unsuccessful fit, likely lens: 20.
    \end{enumerate}
\end{enumerate}

\section{Modelling}
\label{sec:Modelling}
\subsection{Modelling overview}
We employed \texttt{PyAutoLens} (\citealp{Nightingale_2021}) to test whether the observed configurations could be reproduced by realistic mass models and background sources. This step is critical in discriminating between actual strong lenses and lens-like contaminants, such as mergers, ring galaxies, or random alignment of galaxies.By fitting both the lens light and the lensed source light, the modelling pipeline quantitatively assesses the plausibility of each candidate as a strong lens. Only systems for which physically consistent lens models could be recovered were retained as confirmed lenses, reducing false positives and enabling further scientific analysis.

We perform automated strong lens modelling of all candidates with grades above 1.5 using the \Euclid Strong Lens Modelling Pipeline\footnote{\href{https://github.com/Jammy2211/euclid\_strong\_lens\_modeling\_pipeline}{github.com/Jammy2211/euclid\_strong\_lens\_modeling\_pipeline}}, which is built on the open-source lens modelling software \texttt{PyAutoLens}\footnote{\href{https://github.com/Jammy2211/PyAutoLens}{github.com/Jammy2211/PyAutoLens}}
 (\citealp{Nightingale_2021}). The pipeline configuration follows that introduced by SLDE A, designed for high-throughput analysis of \Euclid-quality data.

The lens mass distribution is modelled as a singular isothermal ellipsoid (SIE),
\begin{equation}
\label{eqn:SPLEkap}
\kappa (\xi) = \bigg(\frac{1}{1 + q^{\rm mass}}\bigg) \frac{\theta^{\rm mass}_{\rm E}}{\xi}\;, 
\end{equation}
with
\begin{equation}
    \xi = \sqrt{x^2+\frac{y^2}{(q^{\rm mass})^2}}\;,
\end{equation}
where $\kappa$ denotes the dimensionless (normalized) surface mass density, defined as $\kappa = \Sigma / \Sigma_{\mathrm{crit}}$, with $\Sigma$ the projected mass density and $\Sigma_{\mathrm{crit}}$ the critical surface mass density. 
Here, $\theta^{\rm mass}_{\rm E}$ is the Einstein radius, $q^{\rm mass}$ is the axis ratio, and $\xi$ is the elliptical coordinate from the centre in the image plane. The rotated and shifted Cartesian coordinates, $x$ and $y$, are aligned on the image centre and rotated to the chosen orientation. Deflection angles are calculated following \citet{kormann_isothermal_1994}, as implemented in \texttt{PyAutoLens}. External shear is included, parametrized by $(\gamma_1^{\rm ext}, \gamma_2^{\rm ext})$, with shear magnitude and orientation defined as
\begin{equation}
    \label{eq:shear}
    \gamma^{\rm ext} = \sqrt{\gamma_{\rm 1}^{\rm ext^{2}}+\gamma_{\rm 2}^{\rm ext^{2}}}, \quad
    \tan{2\phi^{\rm ext}} = \frac{\gamma_{\rm 2}^{\rm ext}}{\gamma_{\rm 1}^{\rm ext}}\;.
\end{equation}

The lens light is first fitted and subtracted using a multi-Gaussian expansion \citep[MGE;][]{Cappellari2002,he2024mge}, and PSF blurring is accounted for in all model stages. The mass model is used to ray-trace image pixels to the source plane, where the source light is reconstructed on a pixelised mesh. The pipeline proceeds through five sequential model-fitting stages of increasing complexity. Initial stages employ parametric (MGE) source models to ensure rapid and stable convergence, while later stages adopt adaptive mesh reconstructions using Voronoi or Delaunay tessellations for greater flexibility. This combination of steps allows robust fitting of both smooth and complex lensed structures. 

\newpage

The pipeline is largely automated and designed to handle large candidate samples, although some steps, such as preparing masks, still require manual intervention. Further details on the \texttt{PyAutoLens} framework can be found in \citet{nightingaleScanningDarkMatter2024} and \citet{he2024mge}. The outline of this pipeline can be seen in Table 2 of \citet{Q1-SP048}.

\begin{figure*}
\centering
\includegraphics[width=\textwidth]{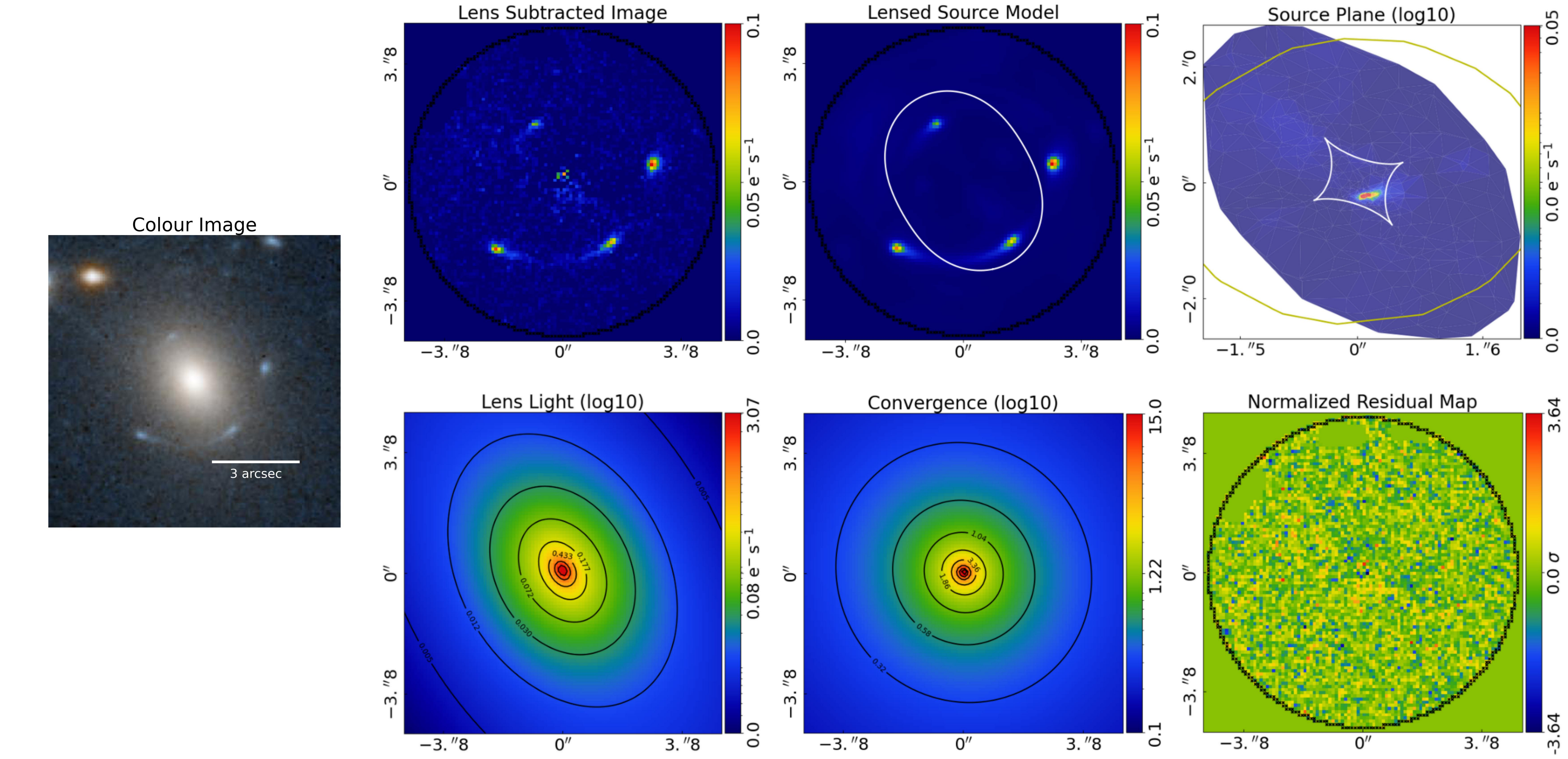}
\includegraphics[width=\textwidth]{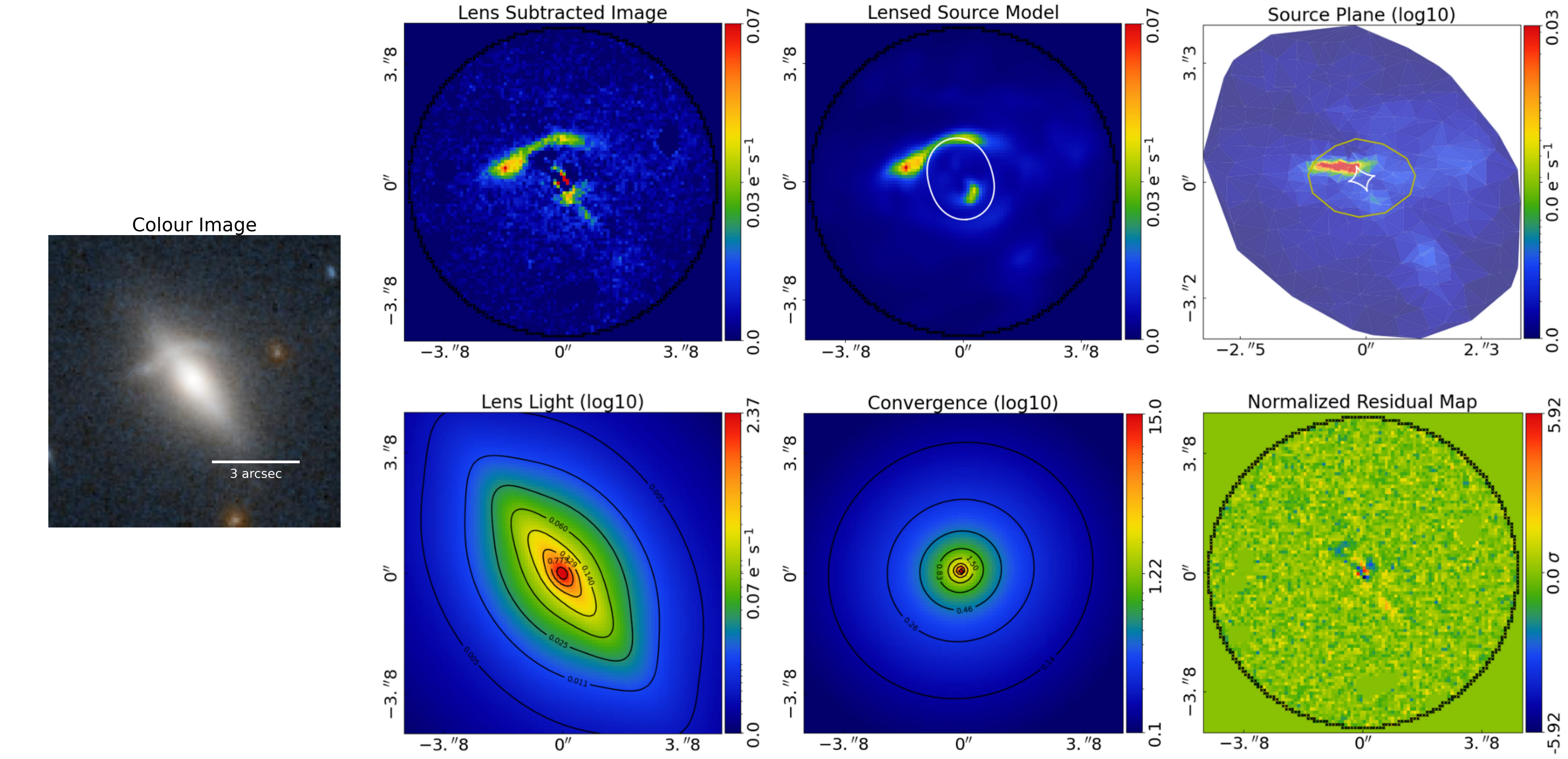}
\caption{Examples of lens models produced by the \Euclid Strong Lens Modelling Pipeline. 
From left to right, the first panel shows the \IE+\JE colour image with MTF scaling. 
The top row displays the lens-subtracted image, the lensed source model, and the source-plane reconstruction on a Delaunay mesh. 
The bottom row shows the lens light model, the SIE convergence $\kappa$, and the normalised residuals of the fit. 
White and yellow lines indicate the tangential critical line/caustic and the radial caustic, respectively. 
The black circle marks the circular $\ang{;;5}$ mask that was applied.}
\label{fig:lens_models}
\end{figure*}

\subsection{Modelling results}
The \Euclid Strong Lens Modelling Pipeline was applied to 72 lens candidates that had received a high expert score (Grade A or B, corresponding to a reweighted average of above 2 and 1.5, respectively). The first step determined whether the automated modelling was successful, primarily by assessing how well the model reproduced the observed lensed source emission. We examined the model’s critical curves, the reconstructed source plane for physical plausibility, and the normalised residual map for flatness. 

A successful model does not necessarily confirm the object as a strong lens; it simply indicates that the model accurately fits the observed image. For example, if the emission in the image plane appears as a single image with no counter-image and the model reproduces this configuration, the fit is considered successful. However, such cases require further interpretation: the analysis included checking whether the source lies within the caustic structure of the lens model. In particular, it was investigated whether the source is entirely outside the caustic (resulting in no multiple images) or lies within a naked cusp caustic, which can also produce configurations without a counter-image. This distinction is important for understanding whether a successful model represents a genuine strong lens or simply a plausible fit to the observed light distribution. Out of the candidates modelled, 41 received a successful fit and are considered likely lenses. Two examples of successful lens models are displayed in Fig.~\ref{fig:lens_models}, and \autoref{tab:lens_table} summarises key information for each lens candidate.

\section{\label{sec:Redshifts}Redshift of EUCL\,J180412.90+655751.1 and EUCL\,J180727.61+653803.6}
\subsection{HET data}
We obtained spectra for EUCL\,J180412.90+655751.1 and EUCL\,J180727.61+653803.6 using the Visible Integral-field Replicable Unit Spectrograph (VIRUS; \citealt{HILL2006378, Hill_2021}), the main wide-field spectrograph on the Hobby-Eberly Telescope (HET) at McDonald Observatory. This is part of the Texas \Euclid Search for Ly~$\alpha$ survey (TESLA, \citealp{ChavezOrtiz2023}) which is an unbiased spectroscopic survey covering approximately $10\deg^2$. VIRUS is a massive integral field spectrograph designed for large-area spectroscopic surveys, consisting of 78 Integral Field Units (IFUs) with 448 fibres each. It covers a 22\arcmin\ diameter field of view. The instrument delivers spectral coverage from approximately 3500 to 5500\,\AA\ at a resolution of \(R \simeq 800\) at 4500\,\AA, making it well suited for low-redshift galaxy spectroscopy and emission-line identification. The spectra are shown in Figs.~\ref{fig:spectra1} and \ref{fig:spectra2}.
\subsection{Redshift measurement}
To determine the redshift and aperture-averaged stellar velocity, the publicly available penalised PiXel Fitting (PPXF, \citealp{Cappellari2017, Cappellari_2023}) method was used. For EUCL\,J180412.90+655751.1 the redshift of $z=0.2872\pm0.0001$, with a velocity dispersion of $\sigma_{\text{ap}}=(246\pm29)\,\rm{km}\,\rm{s}^{-1}$ is obtained. This is in good agreement with the photometric redshift estimation $z_{\text{phot}}=0.26$. For EUCL\,J180727.61+653803.6 we obtain a redshift measurement of $z=0.2636\pm0.0001$ with $\sigma_{\text{ap}}=(259\pm15)\,\rm{km}\,\rm{s}^{-1}$ and a photometric redshift of $z_{\text{phot}}=0.29$. The whole sample of spectra and redshifts for the early-type galaxies (ETGs) within the \Euclid footprint, along with a detailed description of the methodology, will be presented in Balzer et al. (in prep.).

\section{\label{sec:Results}Results and summary}

\subsection{Lens catalogue and derived data products}
The full sample of 72 strong lens candidates discovered with the methods described in Sect.~\ref{sc:Method} is presented in \autoref{tab:lens_table}. The table lists the object ID (column 1), the right ascension and declination coordinates (columns 2 and 3), and the photometric redshift $z_{\mathrm{phot}}$ (column 4) from the MER catalogue \citep{Q1-TP005}. Where available, spectroscopic redshifts $z_{\mathrm{spec}}$ are included (column 5), based on external databases and VIRUS observations. Each candidate is further assigned an expert visual inspection (VI) score (column 6) as described in Sect.~\ref{sec:VIExpert}, and the $I_E$ magnitude (column 7) and Sérsic index $n_{\text{sersic}}$ (column 8) from \citet{Q1-SP040}, the latter serving as a proxy for morphology. Stellar masses (column 9) are taken from \citet{Q1-TP005}. The Einstein radius $\theta_{\mathrm{E}}$ (column 10) is derived from lens modelling when successful, and the comments column (column 11) highlights noteworthy features of individual systems, discussed further in Sect.~\ref{sec:ind_obj}. The table and data products are available on Zenodo.\footnote{\url{https://doi.org/10.5281/zenodo.19260200}}

\subsection{Difference to SLDE A}
The first visual inspection, described in \citep{Q1-SP066}, included the removal of foreground galaxy light to reveal buried structures. Features such as arcs become visible after subtracting MGE or S\'ersic models, since arcs are not radially symmetric. Moreover, the radial profiles of MGE and S\'ersic models are smooth, so even Einstein rings can be effectively captured. This subtraction aids the identification of counter-images, thereby improving the efficiency of grading and strong lens identification.

\begin{figure}[htbp!]
\centering
\includegraphics[angle=0,width=1.0\hsize]{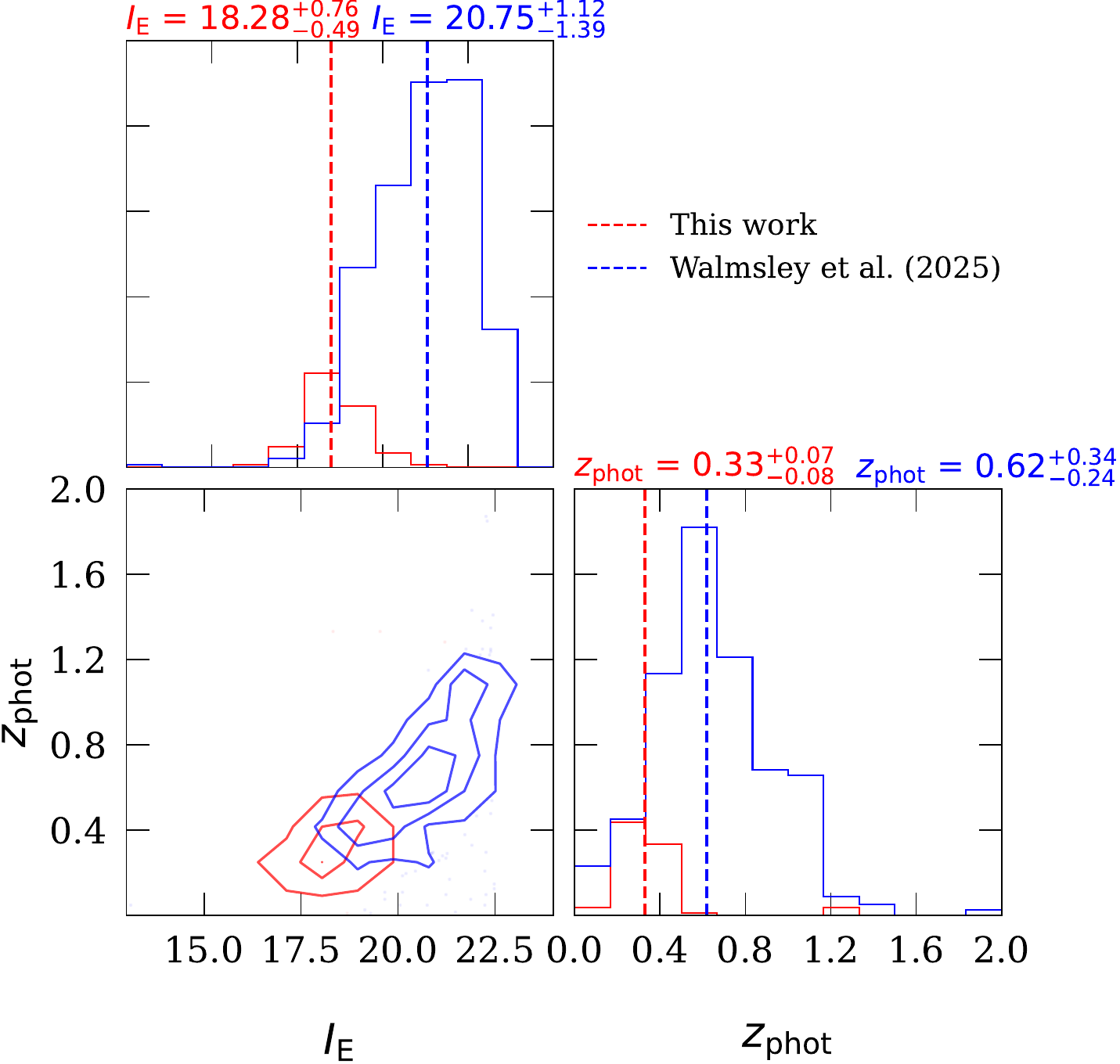}
\caption{Distribution of both strong lens samples with grade A+B with their corresponding photometric redshift and brightness within the Kron Aperture. The blue histogram shows the distribution for the SLDE A catalogue, while the red histogram represents the distribution from this work. The dashed lines indicate the median values for each sample.}
\label{fig:corner_z_mag}
\end{figure}

This work builds on the previous SLDE study by using a retrained \texttt{Zoobot} model, using the SLDE A positive test set. The exclusion of \textit{Gaia} objects in SLDE A introduced a bias against low-redshift and bright sources.
All the 72 newly discovered objects possess a {\tt gaia\_id}. Two  pieces of evidence support our claim that we found a distinct lens population that was not found by SLDE A. The redshift and their corresponding luminosity distributions within the Kron aperture from the MER catalogue for both samples are shown in Fig. \ref{fig:corner_z_mag}. The two samples probe a different distribution of parameter space within magnitude \IE and photometric redshift $z_{\text{phot}}$. However, this should be taken with some caution since photometric redshifts are not as reliable as spectroscopic redshifts.

\begin{figure}[htbp!]
\centering
\includegraphics[angle=0,width=1.0\hsize]{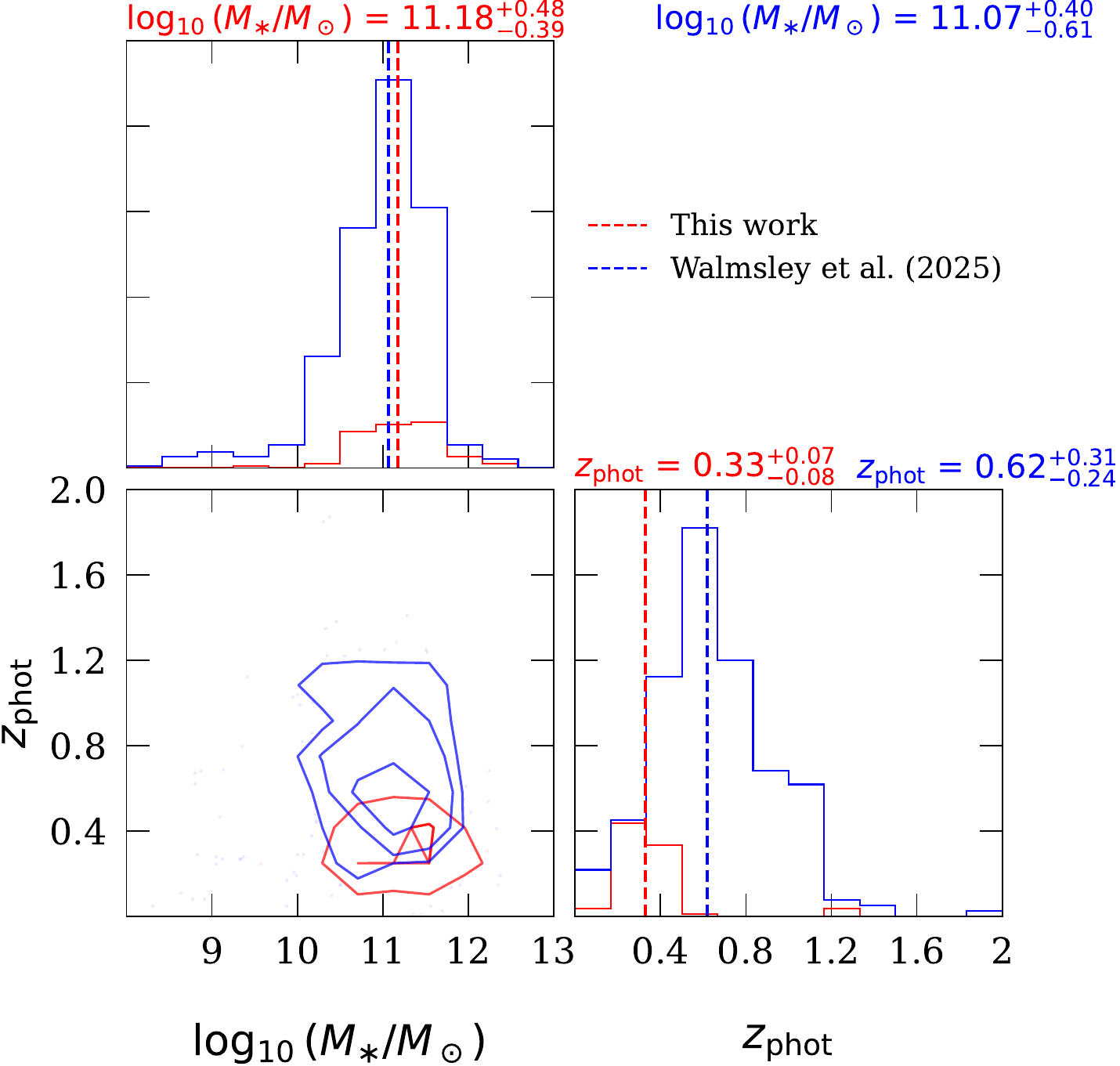}
\caption{Distribution of both strong lens samples with grade A+B in the context of photometric redshift and stellar mass. The blue histogram shows the distribution for the SLDE A catalogue, while the red histogram represents the distribution from this work. The dashed lines indicate the median values for each sample. We note that the displayed one sigma lines are affected by our small sample size but the offset between the measurements are clearly visible and real.}
\label{fig:corner_z_stellarmass}
\end{figure}

\begin{figure}[htbp!]
\centering
\includegraphics[angle=0,width=1.0\hsize]{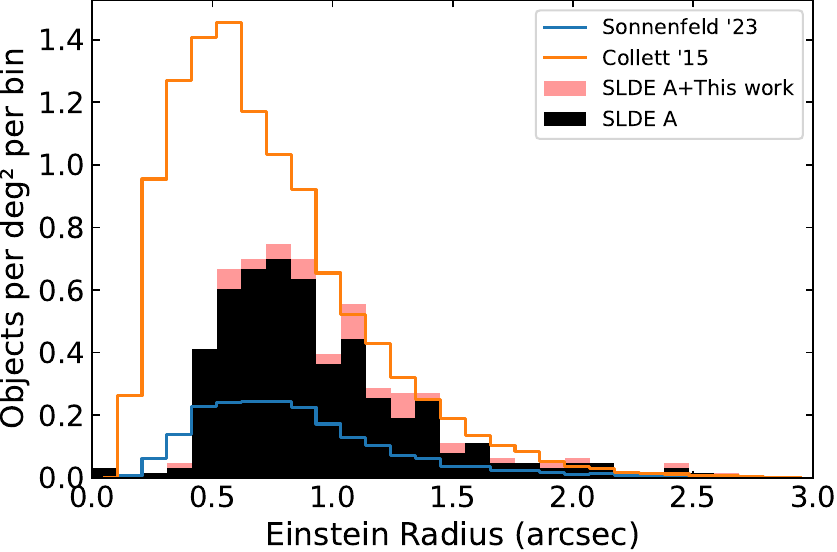}
\caption{Distribution of modelled Einstein radii. Histograms depicting the distribution of succesful strong lens models of the lens population found in SLDE (black) and added with this work (red). Overlayed are the expected Einstein radii of lenses in the forecast made by \citet{Collett_2015} and \citet{Sonnenfeld_2023}.}
\label{fig:einstein radii}
\end{figure}

As shown in Fig. \ref{fig:corner_z_stellarmass}, there is no indication of a selection bias against stellar mass when compared to this work and SLDE A. Due to their lower redshifts, the strong lenses in this study probe more central regions of the galaxies in physical units (kpc) compared to SLDE A, despite having a similar Einstein radius distribution (Fig. \ref{fig:einstein radii}).

After successfully modelling 41 strong lenses, the number density distribution of Einstein radii per bin is compared to the forecast of \citet{Collett_2015} and to the lens sample presented in \citet{Sonnenfeld_2023}. Since both SLDE A and this work focus on Q1, the increase in the number of detected objects relative to SLDE A is indicated by the light red colour.

We emphasize that \citet{Sonnenfeld_2023} is not a forecast of the strong-lensing population. Their lens sample is constructed under significantly stricter selection criteria, including a truncation of the lens population at $z_\mathrm{lens} < 0.7$, a source population limited to $z_\mathrm{source} < 2.5$, and a more restrictive definition of what constitutes a strong lens. These choices naturally lead to a substantially lower number of lenses compared to the forecast of \citet{Collett_2015}, and the two distributions should therefore not be interpreted as directly comparable predictions.

\section{\label{sec:Conclusion}Conclusions}
\Euclid is uniquely suited to discovering strong gravitational lenses, combining space-based resolution with wide-area coverage. In this work, we present an extension to the main catalogue of strong lens candidates from the Q1 data release, compiled using a combination of ML, citizen science, expert inspection, and lens modelling which is collectively referred to as the SLDE.

To identify the 497 candidates in SLDE A, five ML models were applied, including \texttt{Zoobot} (\citealp{Q1-SP053}), to approximately one million galaxies in the Q1 dataset. These models were trained using a combination of simulated lenses and a small set of spectroscopically confirmed systems discovered within \Euclid (\citealp{Q1-SP052}).The final catalogue contains 72 strong lens candidates that are distinct from the previous Q1 search (SLDE A--E), of which only three were previously identified in other works.

A critical insight gained during this process was the identification of a selection bias in the candidates. As shown in Figs. \ref{fig:corner_z_mag} and \ref{fig:einstein radii}, the distribution of lens candidates reveals a clear underrepresentation of high-surface-brightness, low-redshift systems, as well as intermediate-Einstein-radius galaxies acting as lenses. This bias did not arise from the performance of the ML algorithms, but from the selection cuts applied prior to candidate evaluation, which effectively excluded such systems from the input sample. Consequently, retraining the models on the discovered lenses could propagate this bias, since the algorithms would have had no opportunity to learn or identify these otherwise highly detectable lenses. Since only a subset of the highest-graded objects (by ML) is inspected by humans, it is crucial to eliminate any systematic biases against specific populations of lens galaxies. It also has to be noted that the predicted 
(\citealp{Collett_2015}) population of small Einstein radii could not be found.

This work, along with \citet{Q1-SP066}, presents the first dedicated search for low-redshift strong lenses based solely on photometric data. The strategy of targeting large, bright galaxies and subtracting their light has proven effective in revealing central features such as multiple nuclei or strongly lensed arcs. 

With the previous catalogue comprising 497 candidates and an additional 72 newly identified lenses from \Euclid, the forecast of $10^5$ strong lenses in the wide survey is still accurate. These forecasts may increase further through the identification of additional systematic biases or improvements in ML methodologies \citep{Q1-SP085}. Notably, the annotations from both catalogues are already being utilised to retrain and enhance ML models (Vincken et al. in prep and Xu et al. in prep).

We obtained spectra for EUCL\,J180727.61+653803.6 and EUCL\,J180412.90+655751.1, measuring lens galaxy redshifts of $z_{\rm spec}=0.2872 \pm 0.0001$ and $z_{\rm spec}=0.2636 \pm 0.0001$, respectively. As detailed in Sect.~\ref{sc:pred_zs}, we infer enclosed masses within the Einstein radius of $M(\textless\kern-0.1em\theta_{\rm E}) = 2.45^{+0.61}_{-0.55} \times 10^{11}~M_{\odot}$ and $M(\textless\kern-0.1em\theta_{\rm E}) = 2.97^{+0.36}_{-0.34} \times10^{11}~M_{\odot}$.

The candidate list in this work comprises objects of individual scientific interest. It includes six galaxies featuring a counter-image near the centre, making them ideal for studies of the stellar IMF or supermassive black holes, and six edge-on disc lenses in a variety of configurations, some more favourable than others for disentangling dark matter from baryonic matter and one system initially identified as a DSPL candidate, which is instead interpreted as two merging galaxies being lensed. A highlight of this catalogue is a particular edge-on disc lens, which, according to lens modelling, appears to lens two distinct background sources: one into a fold-arc Einstein ring; and the other into a double image. This rare configuration supports multiple distinct science cases. 

The combined Q1 strong lens catalogue from SLDE A and this work thus marks the beginning of a new era for lensing science. As \Euclid’s coverage grows, so too will its lens discoveries -- soon surpassing all previous searches combined, and opening the door to transformative cosmological and galaxy evolution studies.

%
%

\begin{acknowledgements}
The main work for this paper was completed before the untimely passing of Dr. Stella Seitz, mentor to the first author (LRE), whose thoughtful guidance and passion for science shaped his career and inspired countless generations of students. This was likely one of the last projects to which she contributed, dedicating hours to discussing and classifying individual objects. Her memory will forever inspire us to, in her words, ``do good science''.\\
MF, RS and RB acknowledge support by the Deutsches
Zentrum fur Luft- und Raumfahrt (DLR) grant 50 QE 1101. We appreciate the help of Alexander Fabricius during the visual inspection of \IE cutouts and in the discovery of some of the lenses presented herein. The work of LAM was carried out at the Jet Propulsion Laboratory, California Institute of Technology, under a contract with NASA.

We gratefully acknowledge the Zooniverse team for their invaluable work on the host platform for Space Warps. 

This project would not be possible without the immense contributions of citizen scientists. Those contributing classifications for this project are listed on \href{https://www.zooniverse.org/projects/aprajita/space-warps-esa-euclid/about/team}{The Team}. We greatly appreciate their committent, enthusiasm, and time.

This publication uses data generated via the Zooniverse.org platform, development of which is funded by generous support, including a Global Impact Award from Google, and by a grant from the Alfred P. Sloan Foundation.

This research makes use of ESA Datalabs (\url{datalabs.esa.int}), an initiative by ESA’s Data Science and Archives Division in the Science and Operations Department, Directorate of Science. This work has made use of data from the
European Space Agency (ESA) mission \textit{Gaia} (\url{https://www.cosmos.esa.int/gaia}), processed by the \textit{Gaia}
Data Processing and Analysis Consortium (DPAC,
\url{https://www.cosmos.esa.int/web/gaia/dpac/consortium}). Funding
for the DPAC has been provided by national institutions, in particular
the institutions participating in the \textit{Gaia} Multilateral
Agreement.
Based on observations obtained with the Hobby--Eberly Telescope (HET), which is a joint project of the University of Texas at Austin, the
Pennsylvania State University, Ludwig-Maximillians-Universit\"at
M\"unchen, and Georg-August Universit\"at G\"ottingen. The HET
is named in honour of its principal benefactors, William P.
Hobby and Robert E. Eberly.
\AckEC  
\AckQone
\end{acknowledgements}

%
\bibliography{Biblio}

%
%

\begin{appendix}

\section{Individual objects with scientific value \label{sec:ind_obj}}

\subsection{DSPL candidate}
\label{sc:DPSL}
\begin{figure*}[htbp!]
\centering
\includegraphics[angle=0,width=1.0\hsize]{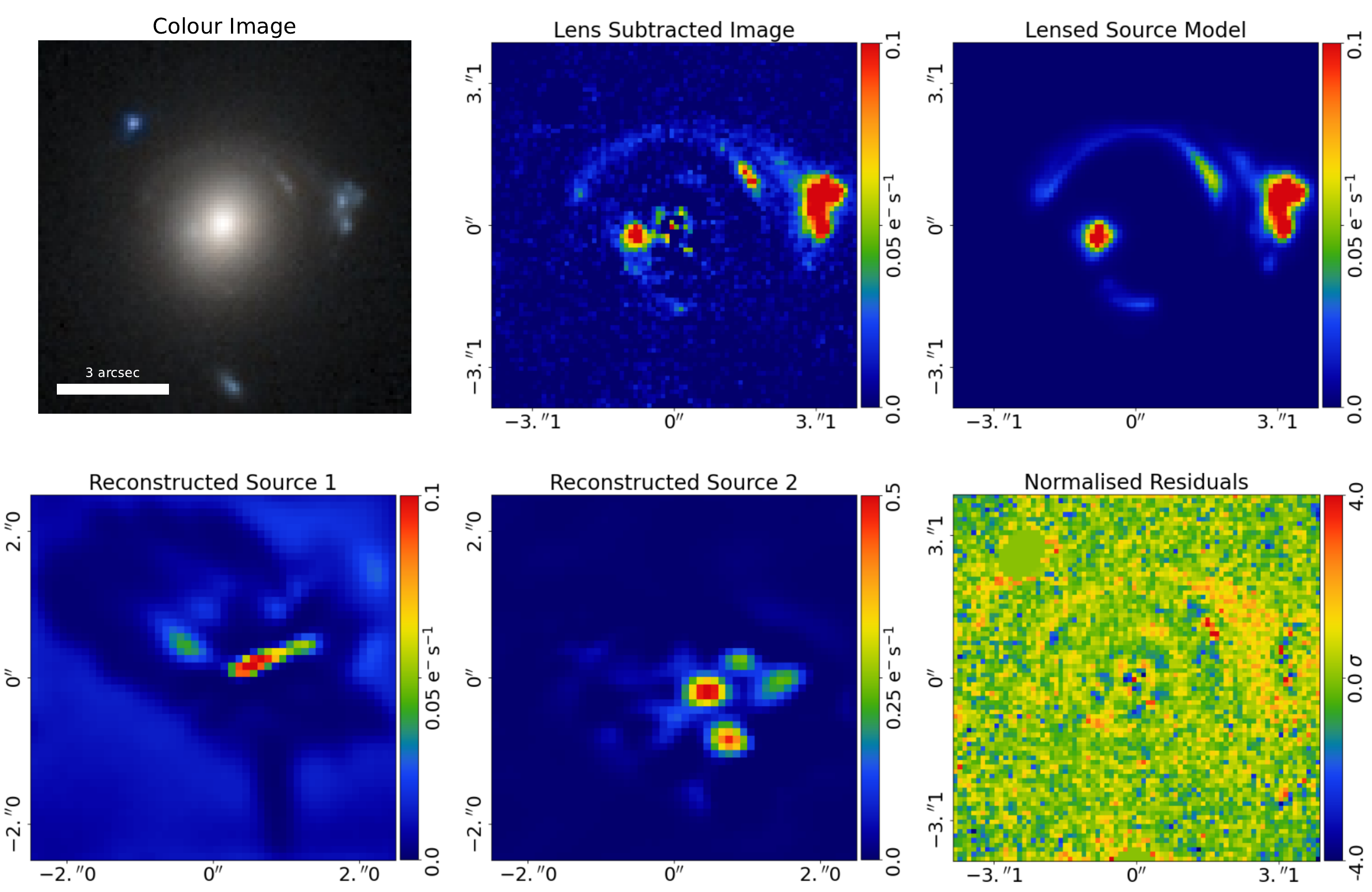}
\caption{\emph{Top row}: top left, colour image of EUCL,J040700.68$-$495323.8 with MTF scaling of \IE+\JE; top middle, lens light-subtracted image; top right, lens model image.
\emph{Bottom row}: bottom left, reconstructed source 1 (producing a big arc on top and smaller arc on the bottom); bottom middle, reconstructed source 2; bottom right, normalised residual image. The critical curves are not well defined in this model and therefore not drawn.}
\label{fig:DSPL}
\end{figure*}

DSPLs are rare configurations where two background sources at different redshifts are lensed by a common foreground galaxy. In these systems, light from each source is lensed, resulting in two sets of lensed images at different distances from the centre of the foreground lens galaxy. These two sets of arcs are particularly useful for constraining the slope of the total mass density profile (e.g. \citealp{Sonnenfeld_2012, Collett_2014, Schuldt_2019}), although measurements of the dark matter distribution and fraction can already be obtained from a single strong lens. After subtracting the lens light using the MGE method \citep{Cappellari2002}, an additional light structure emerged near the centre of EUCL,J040700.68$-$495323.8. This feature could be interpreted as a counter-image belonging to a separate set of arcs. In the colour image of Fig.~\ref{fig:DSPL}, these two arcs exhibit two different colours, motivating the hypothesis that this system could be a DSPL. Since the redshifts of both the lens and the sources are unknown, a full statistical analysis of this system, modeled as a DSPL, is not possible. However, as shown in \citet{Q1-SP054}, we can still assess whether this system is a plausible lens. To do this, we modelled the system using the HERCULENS framework of \citet{Galan2022_herculens} with multiplane lensing capabilities \citep{Enzi2024}. For further details, see SLDE D.

In equation (6) of SLDE D, $\beta$, denoted as the cosmological scaling factor, is defined as the ratio of angular diameter distances between the different redshift planes, 
\begin{equation}
\beta = \frac{D_\mathrm{ls1} D_\mathrm{s2}}{D_\mathrm{s1} D_\mathrm{ls2}},
\end{equation}
where $D$ is the angular diameter distance, and the subscripts l, s1 and s2 denote the lens, source 1 and source 2, respectively. 

Using the model shown in Fig.~\ref{fig:DSPL}, we derive a value of $\beta = 0.986$, which effectively rules out the DSPL hypothesis. This model also fails to reproduce the knots in the image plane of the large arc at the top. We note that modelling the system in the same way as all others with the lens pipeline described in Sect.~\ref{sec:Modelling} also did not yield satisfactory results.

To conclusively distinguish between these scenarios (DSPL or single-plane lens), spectroscopic confirmation would be required. However, this may not be feasible due to the faintness of the arcs, resulting in a long exposure time.

\subsection{Arcs with small projected distance to the centre of mass}
Arcs with small separations to the centre of the lens galaxy are a powerful tool to probe the inner mass distribution. A prime example is the brightest cluster galaxy in Abell 1201 \citep{Edge_2003}, where a faint counter-image very near the core indicated the presence of an additional central mass of order $10^{10} M_{\odot}$, possibly a supermassive black hole \citep{Smith_2017a,Smith_2017b}.
\citet{Nightingale_2023} showed that such a black hole can significantly alter the arc shape in this close vicinity, providing a unique lensing signature of the black hole’s influence. A similar mass measurement of the supermassive black hole from the Cosmic Horseshoe was made by \citet{Schuldt_2019} and \citet{Carlos_Horseshoe2025} by combining strong lens models with dynamical models.

\begin{figure*}[htbp!]
\centering
\includegraphics[angle=0,width=1.0\hsize]{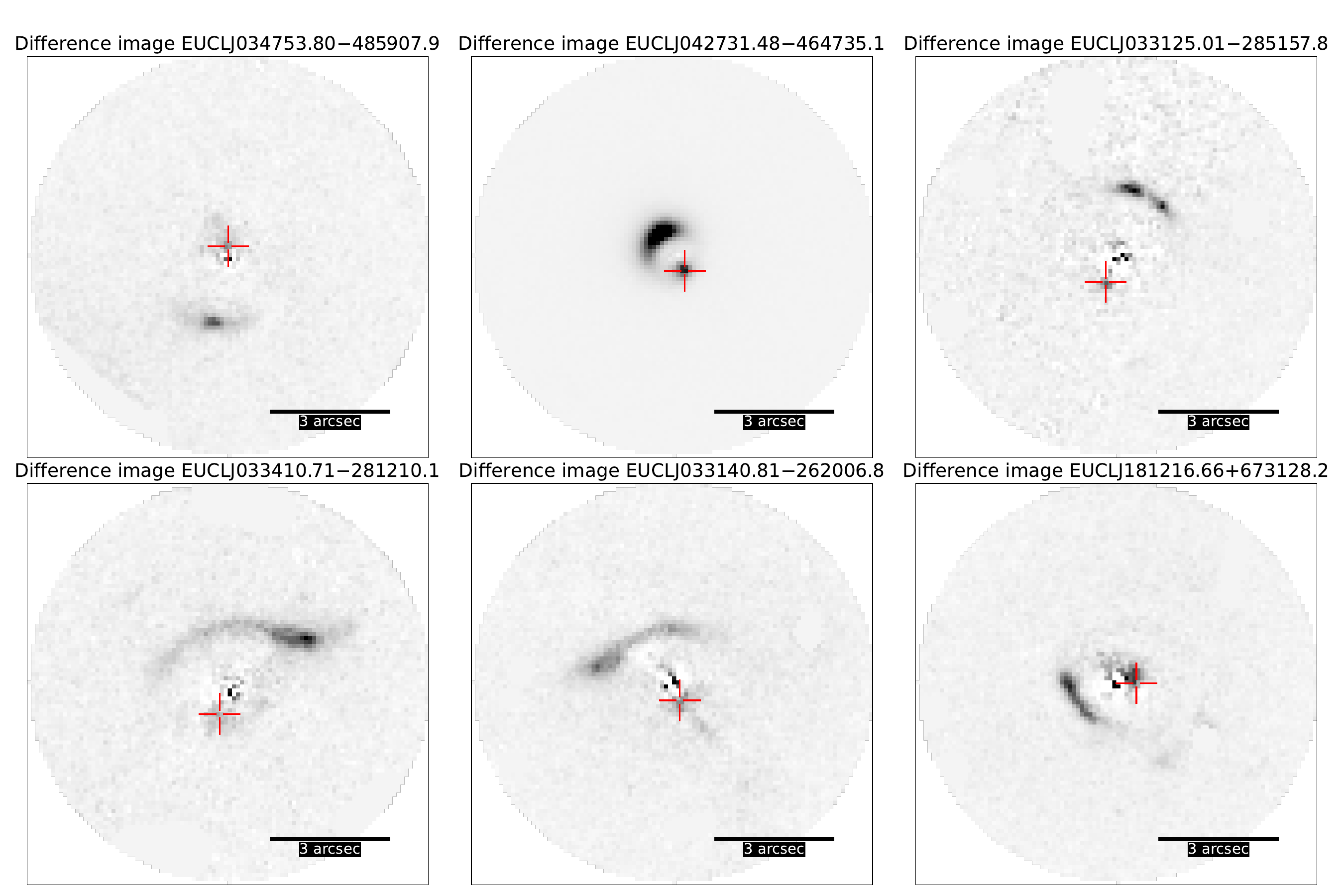}
\caption{Lens light-subtracted candidates showing an arc near the centre of mass. The red cross marks the position of the suspected counterimage, which is in agreement with the lens models.}
\label{fig:smallsep}
\end{figure*}

It has been shown with dynamical Schwarzschild modelling, that within the central roughly 1 kpc of massive early-type galaxies, the mass is dominated by stars rather than dark matter (\citealp{Mehrgan_2024}). This makes small-separation arcs valuable probes for measuring the stellar initial mass function (IMF) and disentangling the contributions of dark matter and stellar mass \citep{ORiordan25, Collett_2018}. Because our sample is at low-redshift, we can combine strong lensing constraints from these arcs with spectroscopic dynamical measurements, leading to improved constraints on both dark matter and the IMF in the innermost regions \citep{Smith_2013, Smith_2015}. Objects falling in this category are EUCL\,J033410.71$-$281210.1, EUCL\,J033125.01$-$285157.8, EUCL\,J033140.81$-$262006.8, EUCL\,J042731.48$-$464735.1, EUCL\,J034753.80$-$485907.9, and EUCL\,J181216.66$+$673128.2. The lens subtracted images are displayed in Fig. \ref{fig:smallsep} and the physical separation is listed in the column \textit{Comments} in \autoref{tab:lens_table}.

\subsection{Edge-on disc lenses}

Edge-on disc galaxies offer a powerful avenue for disentangling the contributions of stellar and dark matter to galaxy mass profiles. In such systems, the highly elliptical projected mass distribution of the luminous matter differs from the rounder projected mass distribution of the dark matter halo, with strong lensing also offering sensitivity to dark matter substructure (\citealp{Hsueh_2016, Hsueh_2017}). This makes them particularly valuable for breaking the classical disc–bulge–halo degeneracy that complicates dynamical studies of galaxy structure. Strong gravitational lensing, when combined with stellar kinematics, provides complementary constraints on the total mass distribution, enabling separate estimates of bulge, disc, and halo contributions. Determining their dark matter content—both on global scales and in their central regions—as well as their total mass density slope, will provide valuable insights into the physical mechanisms linking late- and early-type galaxies across different masses and redshifts (e.g., \citealp{Posti_2019}; \citealp{Tortora_2019}). In particular, since most of these lenses are expected to be very massive disc galaxies, such analyses can help constrain the golden mass (a characteristic mass scale around $5 \times 10^{10} \, \rm M_{\odot}$ in stellar mass), which corresponds to a break in the stellar-to-halo mass relation and a minimum in the central dark matter fraction, offering a way to interpret this transition within a cosmological framework (\citealp{Tortora_2025}).

Current samples of edge-on disc lenses remain limited. The SWELLS survey \citep{Treu_2011}, with approximately 20 edge-on lens candidates at a median redshift of $z = 0.13$, suffers from small Einstein radii (around 2 kpc), which restrict sensitivity largely to the bulge mass \citep{Brewer_2012, Brewer_2014}. \citet{Dutton_2011} analysed the disc-dominated lens SDSS~J2141$-$0001 by combining dynamical modelling with strong lensing. In this configuration the arc is blended with the thin disc. In our sample we have four objects where the arc is blended with the light of the disc. In two cases we have one long arc and a hint of a counter-image after subtracting the light of the lens (EUCL\,J040408.75$-$460534.5, EUCL\,J033140.81$-$262006.8). In the other two cases we have a nearly complete Einstein ring (EUCL\,J035353.76$-$495055.7, shown in different colour schemes in Fig. \ref{fig:1}) and a 4-image configuration (EUCL\,J040936.75$-$483421.1). Similar to the system CXOCY~J2201$-$3201, where a background quasar is lensed into two images outside the plane of the disc (\citealp{Castander_2006, Chen_2013}), we identified two new systems exhibiting a comparable configuration (EUCL\,J033050.17$-$285209.1, EUCL\,J040333.04$-$502315.9). These objects are at intermediate redshift, which will probe the mass with an Einstein radius larger than 2 kpc. Late-type galaxies are underrepresented among known strong lenses. Based on results from the UNIONS survey, it is estimated that only one strong lens candidate is found per $\num{17000}$ edge-on late-type galaxies \citep{barroso2025searchingstronglensinglatetype}. This work extends the number of candidates by six.

\subsection{EUCL\,J035353.76$-$495055.7}
\begin{figure*}[htbp!]
\centering
\includegraphics[angle=0,width=1.0\hsize]{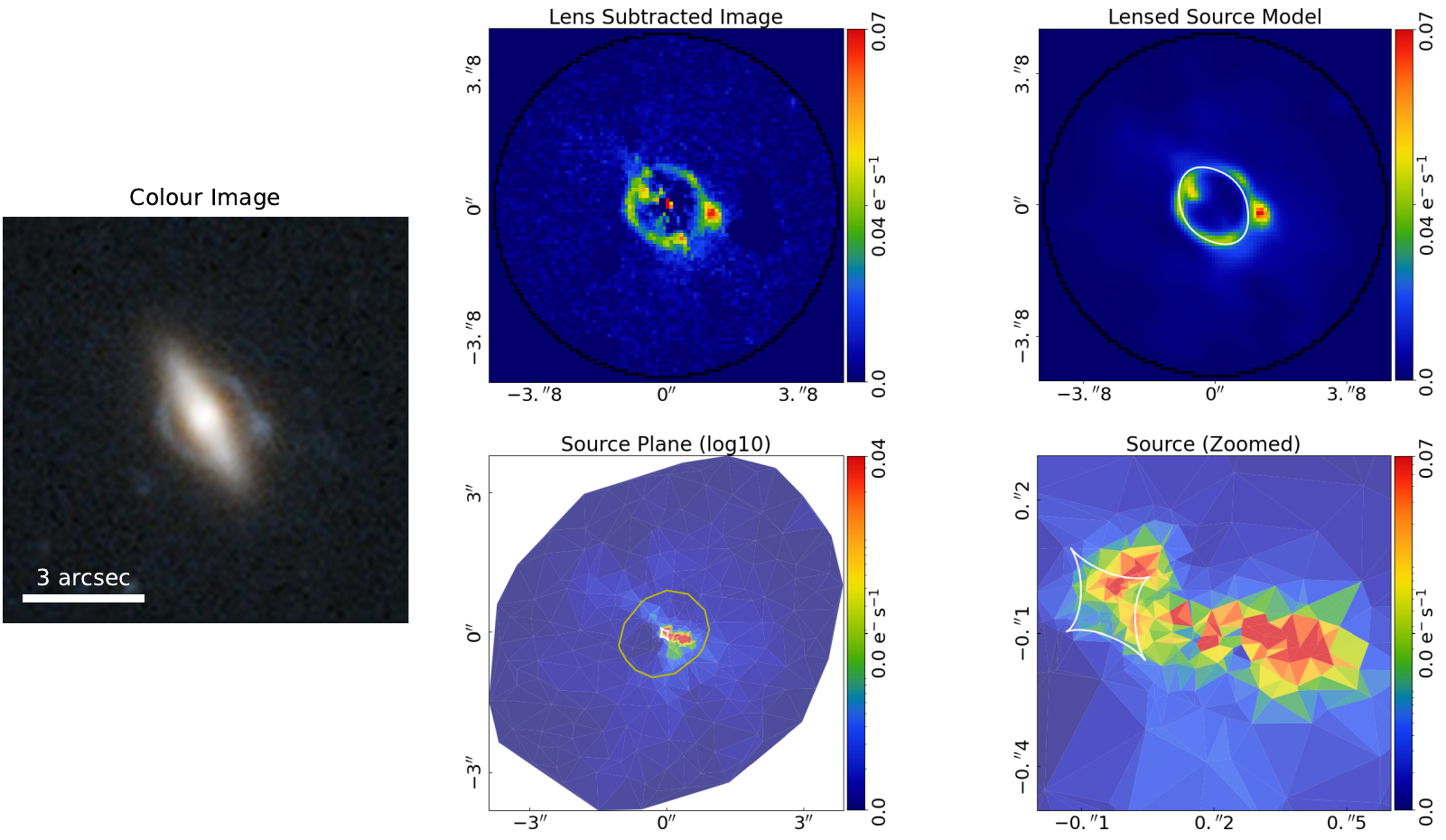}
\caption{Colour image and model of EUCL\,J035353.76$-$495055.7. The colour image is created with an MTF scaling of \IE+\JE. White and yellow lines indicate the tangential critical line/caustic and the radial caustic, respectively. }
\label{fig:merging_model}
\end{figure*}

The lens-light-subtracted image reveals one Einstein ring alongside a pair of images. Lens modelling, including a zoom into the source plane (Fig.~\ref{fig:merging_model}), shows multiple bright structures, some inside and some outside the tangential caustic. This supports an interpretation of either two sources in the same plane that might be merging, or a single irregular source with multiple bright clumps—one forming the fold arc that produces the Einstein ring, and another producing the observed double image. As an edge-on lens system with multiple lensed source components, this object offers compelling science cases for studying dark matter and the stellar IMF, potentially providing improved constraints on the mass profile.

\section{Predicting the redshift of the source \label{sc:pred_zs}}

Since the wavelength range covered by this instrument (HET) does not include any detectable spectral lines from the source, we estimate its redshift by combining the velocity dispersion measurement with the Einstein radius inferred from lens modelling. The Einstein radius used in the SIE profile implemented in \texttt{PyAutoLens} (as defined by Eq.~\ref{eqn:SPLEkap}) is a scaled Einstein radius, which improves numerical efficiency during model fitting. However, the more commonly used definition in the literature refers to the effective Einstein radius, denoted as \(\theta_{\mathrm{E,eff}}\), which is defined via the area \(A\) enclosed by the tangential critical curve

\begin{equation}
\theta_{\mathrm{E,eff}} = \sqrt{\frac{A}{\pi}} \; .
\end{equation}
Throughout this work, whenever we refer to the Einstein radius, we mean this definition. To compare lensing and dynamical mass estimates, we adopt a 3D singular isothermal sphere (SIS) density profile:

\begin{equation}
\rho(r) = \frac{\sigma^2}{2\pi G r^2} \; ,
\end{equation}
where \(\sigma\) is the 3D velocity dispersion, $r$ is the distance to the centre and \(G\) is the gravitational constant. From lensing, the enclosed mass within the Einstein radius $\theta_{\mathrm{E,eff}}$ using the lens redshift, is

\begin{equation}
M(<\theta_\mathrm{E}) = \pi \theta_{\mathrm{E,eff}}^2 D_{\mathrm{l}}^2 \Sigma_{\mathrm{crit}} \; ,
\end{equation}
with the critical surface density defined as

\begin{equation}
\Sigma_{\mathrm{crit}} = \frac{c^2}{4\pi G} \frac{D_{\mathrm{s}}}{D_{\mathrm{l}} D_{\mathrm{ls}}},
\end{equation}
with $D_{\mathrm{l}}$, $D_{\mathrm{s}}$, and $D_{\mathrm{ls}}$ denoting the angular diameter distances to the lens, to the source, and from the lens to the source, respectively.

The velocity dispersion measured from spectra (e.g., using pPXF) is aperture-weighted, and generally differs from the central model velocity dispersion \(\sigma_0\). To relate them, a projection coefficient \(c_{\rm p}(R)\) must be applied. Following standard derivations \citep{Binney_1982, Eliasdottir_2007, Agnello_2014, Bergamini_2019, Tortora_2009}, the aperture velocity dispersion is given by
\begin{equation}
\begin{aligned}
\sigma_{\mathrm{ap}}^2 &= \frac{4 G}{3 \, L(R)} \Biggl[
\int_0^R R' \, I(R') 
\int_0^{R'} \frac{4 \pi \, \rho(r) \, r^2}{\sqrt{R'^2 - r^2}} \, \mathrm{d}r \, \mathrm{d}R' \\
&- \int_R^\infty R' \, I(R') 
\int_R^{R'} 
\frac{\partial_r \Bigl[ M(r) \, (r^2 - R^2)^{3/2} / r^3 \Bigr]}{\sqrt{R'^2 - r^2}} \, \mathrm{d}r \, \mathrm{d}R'
\Biggr] \\
&\equiv \sigma_0^2 \, c_{\mathrm{p}}^2(R)
\end{aligned}
\end{equation}
with $\partial_r$ denoting the partial derivative with respect to $r$, $I(R)$ the surface brightness profile, $L(R)$ the total luminosity within aperture radius $R$, defined as
\begin{equation}
L(R) = 2\pi \int_0^R R' I(R') \, \mathrm{d}R' \; ,
\end{equation}
and $M(r)$ representing the enclosed mass within 3D radius $r$. For the SIS model, the projection coefficient simplifies to \( c_{\rm p} = 1 \), implying that \( \sigma_{\mathrm{ap}} = \sigma_0 \) (\citealp{Bergamini_2019}). Since strong lensing primarily probes the enclosed mass within the Einstein radius, we adopt the SIS model and use the velocity dispersion measured at \(\theta_\mathrm{E}\) to compute the corresponding dynamical mass. We then compare this to the lensing-inferred mass at \(\theta_\mathrm{E}\), which depends on the source redshift through \(\Sigma_{\mathrm{crit}}\). By requiring both masses to match, we can constrain the redshift of the background source. For EUCL\,J180727.61+653803.6, with a lens redshift of $z_{\rm spec} = 0.2872 \pm 0.0001$ and an enclosed Einstein mass of $M(< \theta_{\mathrm{E,eff}}) = 2.45^{+0.61}_{-0.55} \times 10^{11}~M_\odot$, we find a predicted source redshift of $z_{\rm s} = 1.34^{+27.38}_{-0.57}$. The upper limit of the predicted $z_{\rm s}$ is due to the high uncertainty of $\Delta \sigma_{\mathrm{ap}}/\sigma_{\mathrm{ap}} > 10\%$ and better constraints would definitely improve the estimate. For EUCL\,J180412.90+655751.1, with $z_{\rm spec} = 0.2636 \pm 0.0001$ and $M(< \theta_{\mathrm{E,eff}}) = 2.97^{+0.36}_{-0.34} \times 10^{11}~M_\odot$, we infer $z_{\mathrm{s}} = 1.40^{+2.09}_{-0.46}$. These estimates are obtained by calculating the source redshift at which the Einstein mass matches the SIS prediction based on the measured $\sigma_{\rm ap}$ and the assumption that the Einstein radius of SIS does not deviate much from a SIS. A similar approach was used in \citet{Petrillo_2017} to understand if these lenses were realistic. 
\begin{figure*}[htbp!]
\centering
\includegraphics[angle=0,width=1.0\hsize]{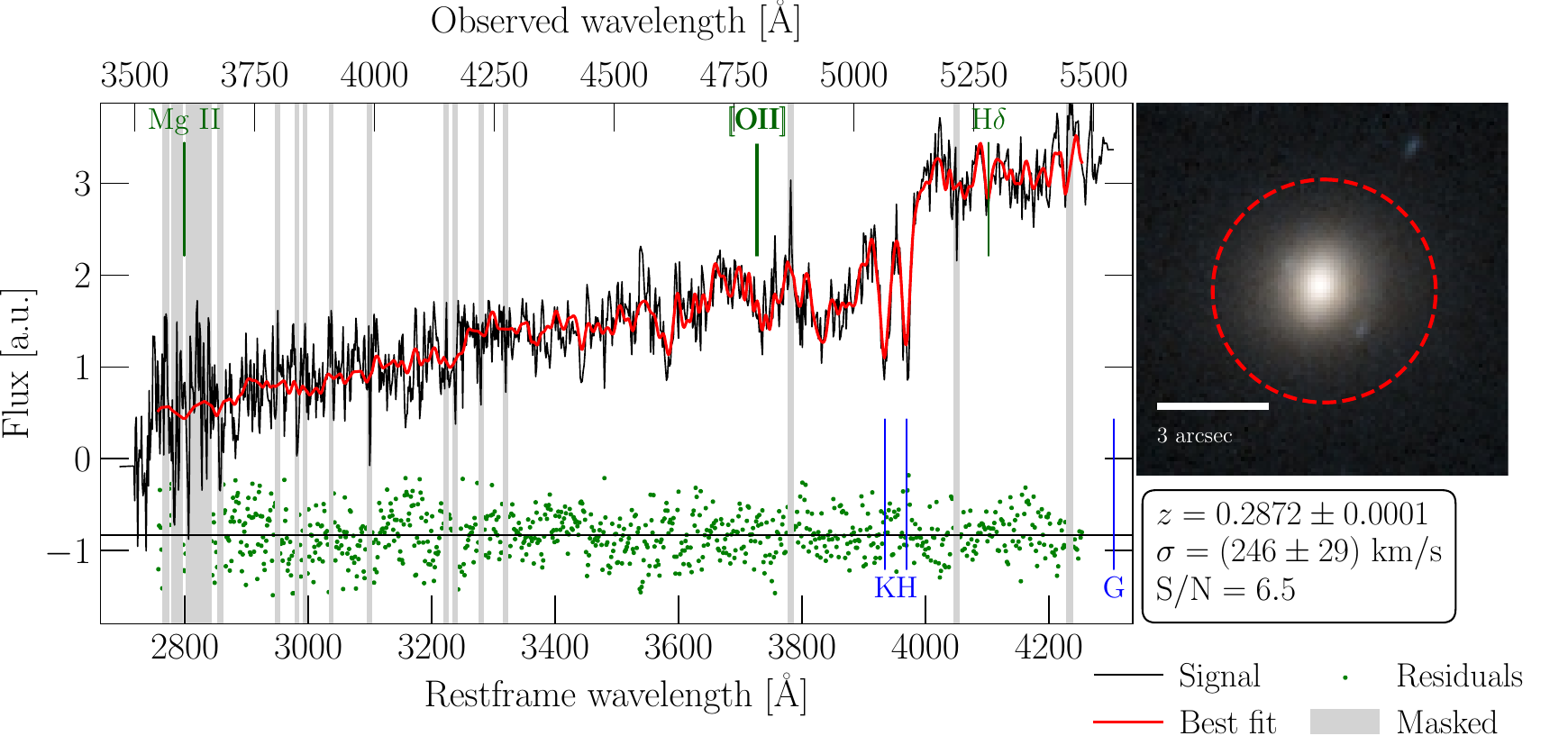}
\caption{Spectrum from EUCL\,J180412.90+655751.1 showing the RGB image of the photometric data from \Euclid on the right with an MTF scaling of \IE+\JE. The dashed red circle with a radius of $\ang{;;3}$ show the aperture which was the size in which the VIRUS IFS data was extracted. On the left side the spectrum is displayed in black along with the fit obtained by \texttt{PPFX} in red. Prominent absorption and emission lines are also marked. }
\label{fig:spectra1}
\end{figure*}
\begin{figure*}[htbp!]
\centering
\includegraphics[angle=0,width=1.0\hsize]{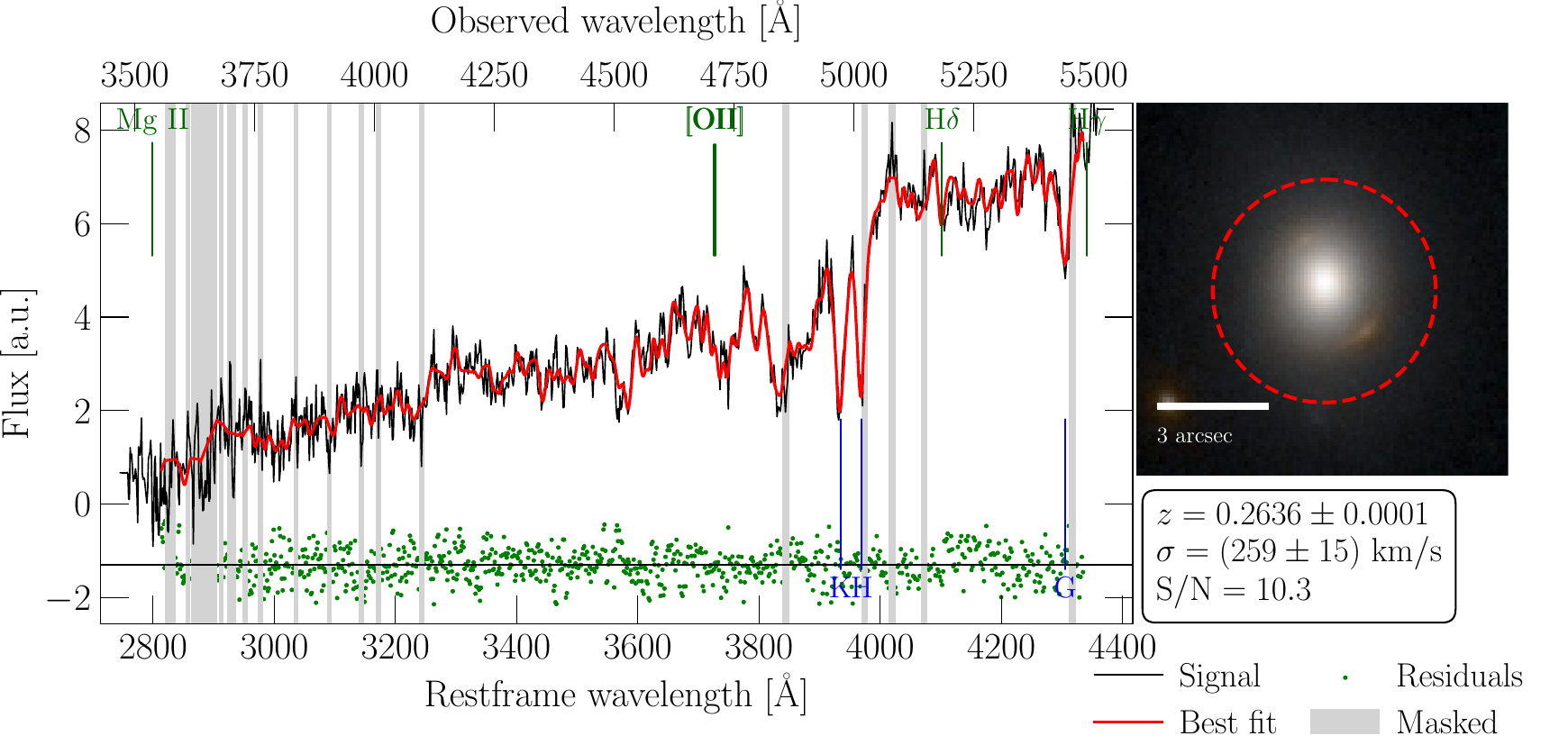}
\caption{Same as Fig.~\ref{fig:spectra1} for EUCL\,J180727.61+653803.6.}
\label{fig:spectra2}
\end{figure*}

\renewcommand{\dblfloatpagefraction}{0.5}

\section{Catalogue of Lens Candidate Properties}
The catalogue of 72 lens candidates, including their derived properties, is presented in this section.
{\sisetup{table-number-alignment = center, round-mode=places, round-precision=6}
\sisetup{
    group-separator = {},
    input-ignore = {},
    detect-weight=true,
    detect-family=true,
    tight-spacing = true,
}
\let\oldpagestyle\pagestyle

\clearpage
\pagestyle{nofooter}

\begin{landscape}
\onecolumn
\setlength{\LTleft}{-200pt}   
\setlength{\LTright}{\fill} 

\begin{longtable}{cSScccccccp{4cm}}
\caption{Lens candidate properties. The full sample of 72 strong lens candidates discovered with the methods described in Sect.~\ref{sc:Method}. (1) The object ID of the candidate; (2) lens galaxy right ascension; (3) lens galaxy declination; (4) photometric redshift of lens galaxy; (5) spectroscopic redshift of lens galaxy: with ``a'' being DESI\_DR1 (\citealp{desicollaboration2025datarelease1dark}), ``b'' being OzDES (\citealp{Lidman_2020}), ``c'' being PRIMUS (\citealp{PRIMUS1},\citealp{PRIMUS1}), and ``d'' being: NEPs (\citealp{ChavezOrtiz2023}, Balzer in prep.); (6) expert visual inspection score; (7) magnitude (\citealp{Q1-SP040}); (8) Sersic index $n_{\rm sersic}$ (\citealp{Q1-SP040}); (9) stellar mass (obtained through $z_{\mathrm{phot}}$ and a Chabrier IMF \citealp{Q1-TP005}); (10) Einstein radius; (11) additional comments.}\label{tab:lens_table} \\ 
\hline
ID & \multicolumn{1}{c}{RA [$\deg$]} & \multicolumn{1}{c}{Dec [$\deg$]} & 
\multicolumn{1}{c}{$z_{\mathrm{phot}}$} & \multicolumn{1}{c}{$z_{\mathrm{spec}}$} & 
\multicolumn{1}{c}{VI score} & \multicolumn{1}{c}{\IE [mag]} & 
\multicolumn{1}{c}{$n_{\rm sersic}$} & 
\multicolumn{1}{c}{$\log_{10}(M_*/M_{\odot})$} & 
\multicolumn{1}{c}{$\theta_{\mathrm{E}}$ [\arcsec]} & Comments \\
 & \multicolumn{1}{c}{[J2000]} & \multicolumn{1}{c}{[J2000]} & & & & & & & & \\
(1) & \multicolumn{1}{c}{(2)} & \multicolumn{1}{c}{(3)} & \multicolumn{1}{c}{(4)} & \multicolumn{1}{c}{(5)} & \multicolumn{1}{c}{(6)} & \multicolumn{1}{c}{(7)} & \multicolumn{1}{c}{(8)} & \multicolumn{1}{c}{(9)} & \multicolumn{1}{c}{(10)} & \multicolumn{1}{c}{(11)}\\
\hline
\endfirsthead
\newpage
\multicolumn{11}{c}{{\tablename\ \thetable{} -- continued from previous page}} \\
\hline
ID & \multicolumn{1}{c}{RA [$\deg$]} & \multicolumn{1}{c}{Dec [$\deg$]} & 
\multicolumn{1}{c}{$z_{\mathrm{phot}}$} & \multicolumn{1}{c}{$z_{\mathrm{spec}}$} & 
\multicolumn{1}{c}{VI score} & \multicolumn{1}{c}{\IE [mag]} & 
\multicolumn{1}{c}{$n_{\rm sersic}$} & 
\multicolumn{1}{c}{$\log_{10}(M_*/M_{\odot})$} & 
\multicolumn{1}{c}{$\theta_{\mathrm{E}}$ [\arcsec]} & Comments  \\
\hline
\endhead

\hline
& & & & & & & & & &\\[-10pt]
\multicolumn{11}{r}{Continued on next page} \\
\endfoot

\hline
\endlastfoot
EUCL\,J033459.29$-$263431.6 & 53.747072 & -26.575460 & 0.36 & 0.346$^b$ & 3.03  & 17.79 & 3.32 & 11.47 & 2.48 & Already found within Euclid Q1 in \citet{Q1-SP057} and in \citet{Jacobs_2019}. \\
EUCL\,J181226.97+671239.9 & 273.112402 & 67.211109 & 0.35 & ... & 3.03 &18.68 & 5.50 & 11.09& 0.75  & ... \\
EUCL\,J041142.54$-$473553.8 & 62.927254 & -47.598300 & 0.33 & ... & 2.98 &  17.85 & 4.57 & 11.48 & 1.13  & Lens isophotes are boxy. \\
EUCL\,J180727.61+653803.6 & 271.865046 & 65.634339 & 0.29 & 0.264$^a$/0.264$^d$ & 2.97 &17.61 & 3.47 & 11.4 & 1.44  & Velocity dispersion $\sigma=(259\pm15)$km\,s$^{-1}$ measured from spectra shown in Fig.~\ref{fig:spectra2}. Source galaxy is red.  \\
EUCL\,J035353.76$-$495055.7 & 58.474032 & -49.848818 & 0.29 & ... & 2.97 &19.01 & 3.0 & 10.88 & 1.05  & Edge-on lens galaxy with two source galaxies. One is lensed in a near complete Einstein ring, the other is doubly lensed. \\
EUCL\,J180512.93+635603.7 & 271.303898 & 63.934380 & 0.39 & ... & 2.96 & 18.1 & 5.50 & 11.57 & 2.65  & Four images \\
EUCL\,J040936.75$-$483421.1 & 62.403125 & -48.572541 & 0.25 & ... & 2.93 &18.19 & 1.59 & 11.0 & 1.09  & The lens galaxy is an edge-on disc lens producing four images. \\
EUCL\,J041644.14$-$480733.1 & 64.183929 & -48.125887 & 0.31 & ... & 2.85 &17.88 & 5.50 & 11.43 & 1.34  & Two bright images. In \HE the arc is significantly larger than in other bands, forming a nearly complete Einstein ring. \\
EUCL\,J174127.24+671517.8 & 265.363505 & 67.254957 & 0.33 & ... & 2.72 &17.83 & 4.76 & 11.38 & ...  & Multiple arcs visible at large distance from the centre, however larger cutouts do not indicate a cluster environment. \\
EUCL\,J175040.08+675456.7 & 267.667012 & 67.915755 & 0.28 & ... & 2.69 &18.52 & 3.04 & 11.69 & 1.04  & Near complete Einstein ring. In \YE the large arc is connected. In \JE and \HE the large arc is split into two separate bright arcs. \\
EUCL\,J040700.68$-$495323.8 & 61.752858 & -49.889960 & 0.26 & ... & 2.66 &17.26 & 4.31 & 10.82 & ... & DSPL candidate and brightest object of the sample. Found already in \citet{Storfer_2024}. \\
EUCL\,J040417.25$-$483113.8 & 61.071905 & -48.520515 & 0.25 & ... & 2.60 &18.36 & 4.96 & 9.57 & 1.08  & ... \\
EUCL\,J175226.01+640221.4 & 268.108393 & 64.039298 & 0.38 & ... & 2.59 &18.07 & 2.23 & 11.9 & ...  & Group-scale lens. \\
EUCL\,J032744.01$-$264747.0 & 51.933381 & -26.796398 & 0.41 & ... & 2.57 &19.78 & 3.72 & 10.67 & 0.69  & ... \\
EUCL\,J033104.06$-$283341.8 & 52.766935 & -28.561636 & 0.28 & ... & 2.56 &17.36 & 3.67 & 11.13 & ...  & ... \\
EUCL\,J033410.71$-$281210.1 & 53.544626 & -28.202830 & 0.38 & 0.289$^b$ & 2.55 &18.26 & 4.42 & 11.53 & 1.33  & Counter-image of the main arc is at a distance of $\sim$\ang[round-mode=places, round-precision=2, round-pad=false]{;;0.8} from the centre of mass. This corresponds to $\sim$3.6\,kpc. Source reconstruction indicates an edge-on disc-like morphology for the source galaxy. Found in \citet{Petrillo_2019}\\
EUCL\,J040328.46$-$475644.6 & 60.868609 & -47.945741 & 0.34 & ... & 2.54 &19.04 & 3.0 & 11.16 & 0.91  & Edge-on lens galaxy \\
EUCL\,J033125.01$-$285157.8 & 52.854217 & -28.866063 & 0.36 & ... & 2.52 &18.6 & 5.06 & 11.08 & 1.25  & A  counter-image is seen at $\sim$\ang[round-mode=places, round-precision=2, round-pad=false]{;;0.85} from the centre of mass, corresponding to $\sim$4.4\,kpc. The larger arc contains two knots that are clearly detected in all infrared filters as well (\YE, \JE, and \HE). \\
EUCL\,J032624.88$-$285246.4 & 51.603669 & -28.879577 & 0.29 & ... & 2.49 &18.8 & 4.74 & 10.94 & ...  & ... \\
EUCL\,J034049.71$-$291543.5 & 55.207162 & -29.262091 & 0.33 & ... & 2.46 &17.46 & 5.50 & 11.76 & ...  &  Group-scale lens\\
EUCL\,J033140.81$-$262006.8 & 52.920071 & -26.335230 & (1.33) & ... & 2.45 &18.29 & 3.71 & (7.87) & 1.18  & One arc is at a distance of $\sim$\ang[round-mode=places, round-precision=2, round-pad=false]{;;0.4} from the centre of mass which corresponds to a physical distance of $\sim$3.4\,kpc. Source reconstruction reveals an edge-on source galaxy. The redshift and the stellar mass are uncertain.\\
EUCL\,J180003.73+633518.0 & 270.015568 & 63.588352 & 0.31 & 0.268$^a$ & 2.35 &18.2 & 3.92 & 12.21 & 1.32  & ... \\
EUCL\,J041357.04$-$452359.7 & 63.487671 & -45.399939 & 0.40 & ... & 2.33 &18.9 & 2.33 & 11.87 & ...  & Group-scale lens. Arc bends in opposite directions because of galaxy in vicinity. \\
EUCL\,J041132.83$-$473621.9 & 62.886812 & -47.606098 & 0.33 & ... & 2.29 &17.98 & 4.47 & 10.71 & 0.54  & ... \\
EUCL\,J034704.32$-$502020.9 & 56.768009 & -50.339161 & 0.39 & ... & 2.23 &18.04 & 5.50 & 11.39 & 1.66  & ... \\
EUCL\,J040333.04$-$502315.9 & 60.887702 & -50.387753 & 0.34 & ... & 2.17 & 19.46 & 5.21 & 10.78  & 0.83  & Edge-on lens galaxy. \\
EUCL\,J035846.20$-$504912.9 & 59.692516 & -50.820253 & 0.38 & ... & 2.16 & 18.61 & 2.28 & 10.81 & 0.89  & ... \\
EUCL\,J040408.75$-$460534.5 & 61.036465 & -46.092942 & 0.25 & ... & 2.15 & 16.65 & 3.29 & 11.5 & 0.98  & Lens galaxy is an edge-on disc galaxy. Lens modeling suggests that the source galaxy is also an edge-on galaxy. \\
EUCL\,J032812.05$-$275056.2 & 52.050244 & -27.848951 & 0.24 & ... & 2.13 & 17.55 & 4.98 & 11.26 & 1.50  & Four knots forming a partial Einstein ring. \\
EUCL\,J032929.39$-$262115.1 & 52.372500 & -26.354217 & 0.31 & ... & 2.12 & 18.21 & 4.84 & 11.33 & ...  & ... \\
EUCL\,J041411.17$-$461227.8 & 63.546556 & -46.207735 & 0.32 & ... & 2.09 & 17.85 & 3.25 & 11.29 & ...  & ... \\
EUCL\,J035023.08$-$510923.0 & 57.596185 & -51.156393 & 0.24 & ... & 2.07 & 17.36 & 4.22 & 10.37 & 1.90  & ... \\
EUCL\,J040631.89$-$495638.0 & 61.632895 & -49.943892 & 0.27 & ... & 2.07 & 17.73 & 4.64 & 10.79 & 0.53  & ... \\
EUCL\,J040727.84$-$465420.0 & 61.866030 & -46.905560 & 1.28 & ... & 2.07 & 21.17 & 5.50 & 11.15 & ...  & Faintest object of the sample. \\
EUCL\,J042052.70$-$470942.9 & 65.219599 & -47.161944 & 0.32 & ... & 2.05 & 17.83 & 5.08 & 12.0 & 1.47  & ... \\
EUCL\,J035111.22$-$482936.2 & 57.796790 & -48.493408 & 0.28 & ... & 2.05 & 17.86 & 4.60 & 11.72 & 1.08  & ... \\
EUCL\,J035524.94$-$471651.7 & 58.853918 & -47.281048 & 0.40 & ... & 2.03 & 18.73 & 4.34 & 10.92 & ...  & ... \\
EUCL\,J033050.17$-$285209.1 & 52.709072 & -28.869195 & 0.26 & ... & 2.03 & 17.79 & 2.40 & 10.79 & 1.27  & Lens galaxy is an edge-on disc. The source is lensed into two images. \\
EUCL\,J042731.48$-$464735.1 & 66.881184 & -46.793100 & 0.23 & ... & 1.99 & 18.04 & 1.66 & 11.63 & 0.54  & Counter-image at \ang[round-mode=places, round-precision=2, round-pad=false]{;;0.64} ($\sim$2.4\,kpc) distance to the centre of mass. The projected size of the source is comparable to the extent of the radial caustic. \\
EUCL\,J042609.10$-$465316.0 & 66.537922 & -46.887798 & 0.20 & ... & 1.99 & 17.58 & 3.84 & 11.38 & 0.78 & ... \\
EUCL\,J042152.08$-$473125.9 & 65.467017 & -47.523869 & 0.43 & ... & 1.97 & 18.84 & 5.50 & 11.24 & 1.10  & ... \\
EUCL\,J033144.66$-$294447.8 & 52.936123 & -29.746622 & 0.28 & ... & 1.96 & 17.86 & 5.50 & 11.18  & ...  & ... \\
EUCL\,J035545.85$-$502833.5 & 58.941060 & -50.475989 & (0.01) & ... & 1.94 & 18.67 & 5.50 & (11.45) & ...  & Redshift and stellar masses are not reliable. \\
EUCL\,J035952.28$-$465754.2 & 59.967840 & -46.965080 & 0.59 & ... & 1.93 & 19.03 & 3.52 & 10.97 & ...  & ... \\
EUCL\,J040026.85$-$492352.9 & 60.111900 & -49.398040 & 0.38 & ... & 1.91 & 18.04 & 4.93 & 11.21 & ...  & ... \\
EUCL\,J175419.78+640204.9 & 268.582455 & 64.034721 & 0.33 & ... & 1.84 & 19.67 & 3.56 & 10.86 & 0.57  & ... \\
EUCL\,J034753.80$-$485907.9 & 56.974203 & -48.985540 & 0.33 & ... & 1.79 & 19.09 & 4.25 & 11.07  & 0.91  & Counter-image at \ang[round-mode=places, round-precision=2, round-pad=false]{;;0.30} distance to the centre which corresponds to $\sim$1.5\,kpc. Source reconstruction suggests an edge-on source galaxy morphology.\\
EUCL\,J034510.54$-$501538.4 & 56.293919 & -50.260669 & 0.41 & ... & 1.79 & 18.99 & 5.50 & 11.0  & 0.63  & Counter-image located at $\sim$\ang[round-mode=places, round-precision=2, round-pad=false]{;;0.2} from centre of mass corresponding to a distance of $\sim$1.2\,kpc. \\
EUCL\,J040347.14$-$483608.4 & 60.946431 & -48.602350 & 0.37 & ... & 1.77 & 19.28 & 2.83 & 10.78 & ...  & ... \\
EUCL\,J181135.91+654538.1 & 272.899655 & 65.760600 & 0.15 & 0.188$^a$ & 1.77 & 17.49 & 1.86 & 12.33 & 0.73  & ... \\
EUCL\,J034400.44$-$492745.7 & 56.001845 & -49.462702 & 0.25 & ... & 1.77 & 17.41 & 4.55 & 11.41 & ...  & ... \\
EUCL\,J034526.11$-$504753.6 & 56.358828 & -50.798244 & ... & ... & 1.75 & & & ...  & & Redshift not available. \\
EUCL\,J034301.95$-$484513.4 & 55.758144 & -48.753731 & 0.41 & ... & 1.74 & 18.64 & 4.54 & 11.38 & ...  & ... \\
EUCL\,J041242.99$-$462543.2 & 63.179152 & -46.428677 & 0.16 & ... & 1.73 & 18.06 & 2.55 & 10.71 & ...  & ... \\
EUCL\,J180412.90+655751.1 & 271.053751 & 65.964202 & 0.26 & 0.288$^a$/0.287$^d$ & 1.73 & 18.63 & 5.47 & 10.89 & 1.24  & ... \\
EUCL\,J041652.71$-$453242.0 & 64.219646 & -45.545024 & 0.35 & ... & 1.68 & 19.60 & 4.52 & 9.79 & ... & ... \\
EUCL\,J032951.40$-$263756.2 & 52.464192 & -26.632281 & (0.01) & ... & 1.67 & 18.55 & 2.30 & 11.57 & 0.51 & ... \\
EUCL\,J040500.78$-$462735.6 & 61.253278 & -46.459916 & (1.33) & ... & 1.67 & 19.53 & 5.50 & (14.41)  & 1.14  & Einstein ring or ring galaxy. Redshift and stellar masses are not reliable. \\
EUCL\,J181216.66+673128.2 & 273.069439 & 67.524512 & 0.21 & ... & 1.68 & 18.22 & 3.25 & 10.9  & 0.81  & Counter-image located at a distance of $\sim$\ang[round-mode=places, round-precision=2, round-pad=false]{;;0.4} ($\sim$1.3\,kpc) from the centre of mass. \\
EUCL\,J173747.36+645853.1 & 264.447345 & 64.981432 & 0.22 & ... & 1.68 & 18.86 & 3.39 & 11.88 & 1.02  & ... \\
EUCL\,J032715.91$-$275427.7 & 51.816315 & -27.907716 & 0.40 & ... & 1.66 & 18.39 & 2.96 & 11.48 & ...  & ... \\
EUCL\,J173131.80+670455.9 & 265.382529 & 67.082215 & 0.15 & ... & 1.63 & 19.10 & 5.50 & 11.46 & ... & ... \\
EUCL\,J042022.04$-$463713.4 & 65.091870 & -46.620408 & 0.28 & ... & 1.63 & 18.38 & 4.31 & 11.08 & ...  & ... \\
EUCL\,J035726.26$-$475336.1 & 59.359450 & -47.893376 & 0.41 & ... & 1.62 & 19.05 & 4.33 & 11.18  & ...  & ... \\
EUCL\,J033859.14$-$281343.8 & 54.746431 & -28.228840 & 0.36 & 0.339$^b$ & 1.61 & 17.85 & 4.52 & 11.72  & ...  & ... \\
EUCL\,J040256.76$-$465635.5 & 60.736508 & -46.943210 & 0.44 & ... & 1.60 & 18.62 & 4.54 & 11.35 & ...  & ... \\
EUCL\,J042404.21$-$465419.6 & 66.017551 & -46.905452 & 0.35 & ... & 1.55 & 19.08 & 4.14 & 10.71 & ...  & ... \\
EUCL\,J032502.74$-$285418.3 & 51.261451 & -28.905102 & 0.22 & ... & 1.53 & 18.49 & 4.01 & 10.81 & ...  & ... \\
EUCL\,J033032.22$-$284809.9 & 52.634290 & -28.802772 & 0.36 & 0.145$^c$ & 1.53 & 18.29 & 4.32 & 10.54 & ...  & ... \\
EUCL\,J040640.23$-$462055.0 & 61.667659 & -46.348627 & 0.38 & ... & 1.52 & 18.28 & 5.5 & 11.48 & ...  & ... \\
EUCL\,J033753.55$-$281828.6 & 54.473165 & -28.307960 & 0.38 & 0.342$^b$ & 1.51 & 17.95 & 5.5 & 11.53 & 2.02  & ... \\
EUCL\,J033428.59$-$275934.7 & 53.619145 & -27.992987 & 0.42 & ... & 1.51 & 19.18 & 5.3 & 11.08 & ... & ... \\ 

\end{longtable}
\end{landscape}
\twocolumn
\pagestyle{otherpage}
}

\end{appendix}

\label{LastPage}

\end{document}